\documentclass{IEEEtran}
\usepackage{amsfonts}
\usepackage{graphicx}
\usepackage{textcomp}
\usepackage{xcolor}
\usepackage{verbatim}
\usepackage{gensymb}
\usepackage{subfigure}
\usepackage{srcltx}
\usepackage{psfrag}
\usepackage{epsfig}
\usepackage{color}
\usepackage[tbtags,sumlimits,nointlimits,reqno]{amsmath}
\usepackage{amssymb}
\usepackage{color}
\usepackage{float}
\usepackage{mathtools, cuted}
\usepackage{commath}
\usepackage[makeroom]{cancel}
\usepackage[normalem]{ulem}
\usepackage{url}
\usepackage{cite}
\usepackage{flushend}

\definecolor{burntorange}{rgb}{0.8, 0.28, 0.0}
\definecolor{myGreen}{rgb}{0.0, 0.5, 0.0}
\definecolor{amber}{rgb}{0.8, 0.28, 0.0}
\definecolor{ceruleanblue}{rgb}{0.16, 0.28, 0.75}
\definecolor{matlabblue}{rgb}{0, 0.4470, 0.7410}
\definecolor{matlaborange}{rgb}{0.8500, 0.3250, 0.0980}
\definecolor{mypurple}{rgb}{0.314, 0.271, 0.588}

\usepackage{overpic}
\usepackage[crop=pdfcrop]{pstool}

\begin{document}
	
	\title{Metasurface Reflector with Real-Time Independent Magnitude and Phase Control}
	\author{Ahmed Z. Ashoor, \IEEEmembership{Member, IEEE} and Shulabh Gupta, \IEEEmembership{Senior Member, IEEE}}
	
	\maketitle
	
\begin{abstract}
	
A novel metasurface unit cell architecture is proposed to enable independent control of the reflection magnitude and phase at a desired operation frequency, while maintaining linear polarization of the incoming fields. The proposed structure is based on a coupled-resonator configuration where a Dipole Ring Resonator (DRR) is loaded with a tunable lumped resistive element (e.g. PIN diode) and Split Ring Resonator (SRR) loaded with a lumped tunable capacitor (e.g. varactor diode), are interleaved. The surface is next operated around one of the coupled resonant frequency, where an independent tuning of the lumped capacitance and resistance elements enable a wide coverage of reflection amplitude-phase, which is significantly larger than what would have been achievable using a single resonator configuration. An insightful equivalent circuit model is further developed for investigating the amplitude-phase characteristics of a uniform surface as a function of variable resistance and capacitance, which is next confirmed using full-wave simulations. Finally, using a variety of full-wave examples, the usefulness of simultaneous and independent amplitude-phase control is demonstrated, including cases of variable pattern gain with beam tilting and multi-beam pattern realization, which otherwise would not be possible using either amplitude or phase control only.
	
\end{abstract}



\section{Introduction}

Electromagnetic (EM) metasurfaces are 2D array sub-wavelength resonators that have recently emerged as a powerful platform for a variety of wave transformations including amplitude and phase control in reflection and transmission, in addition to the polarization manipulation \cite{holloway2012overview,chen2016review,hsiao2017fundamentals,luo2015principles,glybovski2016metasurfaces,arbabi2017fundamental,bukhari2019metasurfaces}. A wide range of applications have been reported to control the reflected EM wave magnitude and phase such as anomalous reflections~\cite{yu2011light}, EM wave absorbers~\cite{kats2012ultra}, high impedance surfaces and artificial magnetic conductors~\cite{sievenpiper1999high}, and beam-forming and beam scanning antennas~\cite{sievenpiper2002tunable,guzman2012electronically,sun2019achieving}, to name a few. A recent trend in metasurface based electromagnetic wave control is to add general space-time modulation of their constitutive parameters for exotic applications such as artificial non-reciprocity and harmonic generation~\cite{caloz2019spacetime, Taravati_Grating,zhang2016advances,bukhari2019metasurfaces}. 

For complete control over the scattered fields at a desired operating frequency $f_0$, metasurfaces unit cells based on sub-wavelength resonators must be capable of providing a full range of amplitude and phase in the desired mode of operation - transmission or reflection, for each orthogonal polarization. For instance, if a reflection mode is desired, the unit cell must \emph{independently} provide $|\Gamma(f_0)|\in [0,~1]$ (perfect absorption to perfect reflection) and $\angle \Gamma(f_0)\in[0,~2\pi]$, i.e. any combination of $\{|\Gamma|, \angle \Gamma\}$. A 2D array of such unit cell elements can thus provide an arbitrary field profile to be generated using a spatially varying complex reflectance. Moreover, if both phase and amplitude of each unit cell can individually be real-time reconfigured on a pixel-by-pixel basis, the resulting metasurface structure may represent an ideal platform for realizing a software controlled \emph{smart reflector}.

While several works have been done to achieve independent control on phase and magnitudes ~\cite{RxInd_Amp_Ph, AmpPhs_HoloMs,liu2014broadband,chen2016geometric}, the majority of them have been restricted to passive metasurfaces only. These techniques typically employ polarization rotation where the unit cell is physically rotated to introduce amplitude modulation of the desired polarization component, while the varying unit cell dimensions are designed for phase control. This introduces spurious cross-polarized components, which may require separate processing to avoid undesired interference with the environment. Moreover, in various practical scenarios, a real-time control on the wave transformation through metasurfaces is highly desirable, such as in application of these surfaces in wireless applications \cite{MS_Wireless}, where the channel characteristics are typically time-varying due to moving objects and people, for instance. In these applications, the metasurface acts as a smart reflector for instance, which can adaptively guide and manipulate the EM waves as the environment dynamically changes. Consequently, the metasurface must be \emph{real-time reconfigurable}.

Lot of work has been done in devising active metasurfaces (i.e.; programmable and tunable metasurfaces), whose reflection/transmission fields can be controlled in real-time~\cite{wan2016field,yang2016programmable,zhu2013active,liaskos2018new,cui2014coding,zhu2014dynamic,zhang2018space,liu2019time}, further enabling software defined control capabilities augmented using Machine Learning (ML) and Artificial Intelligence (AI) techniques~\cite{qian2020deep}. The reconfiguration of the surfaces is based on tunable components and materials which are integrated on a unit cell level. Such techniques are typically based on phase change materials such as graphene, temperature tuning, mechanical tuning or active microwave elements tuning such as PIN (P-type, intrinsic, and N-type material) diodes and varactors~\cite{kats2012ultra,carrasco2013tunable,sievenpiper1999high,sievenpiper2002tunable,guzman2012electronically}. All these works have focussed on tuning \emph{either the magnitude or phase} thereby not utilizing the full capabilities of the metasurface, and thus limited to a small subset of possible wave transformation capabilities. It is clear that a real-time and independent control of magnitude and phase of the scattered fields is an important requirement and a desirable feature in electromagnetic metasurface design to fully exploit the wave control functionalities of general electromagnetic metasurfaces.

In this work, we propose a metasurface unit cell architecture which is capable of providing independent control of the field magnitude and phase at the radio frequencies (RF), in reflection and in real-time without generating any cross-polarized scattered fields. The proposed metasurface is based on a coupled-resonator configuration where each resonator is separately tuned using a varactor and a PIN diode for phase and amplitude control, respectively. The initial idea was proposed in \cite{Ashoor_EuCap} and a similar idea was presented in \cite{ComplexReflectance_Seok} for the optical frequencies in the Terahertz range based on tunable graphene material. Here, a more detailed full-wave demonstration of its phase-amplitude coverage is presented and compared to that of a single resonator. An equivalent circuit model is next presented which is built using single resonators providing further insights into the mechanism for the reflection control. Using several examples, the superiority of the combined amplitude-phase control in achieving a variety of phase transformations over phase-only control is demonstrated using full-wave simulation. 

The paper is organized as follows. Sec.~II motivates the requirements for both amplitude and phase control in achieving general wave transformations and presents the proposed metasurface. Sec.~III shows a step-by-step development of an equivalent circuit model starting from two isolated resonators to a combined coupled-resonator unit cell. Sec.~IV next presents a variety of full-wave examples to demonstrate the utilization of spatially dependent amplitude and phase profiles for far-field beam-forming applications. Sec.~V presents a short discussion on the impact of the unit cell size on the requirements of the lumped capacitance ranges while providing large amplitude-phase coverages at a fixed operating frequency. Finally, conclusions are provided in Sec.~VI.

\section{Proposed Reflection Metasurface}

\begin{figure}[!t]
	\centering
	\begin{overpic}[grid=false, width=0.8\columnwidth]{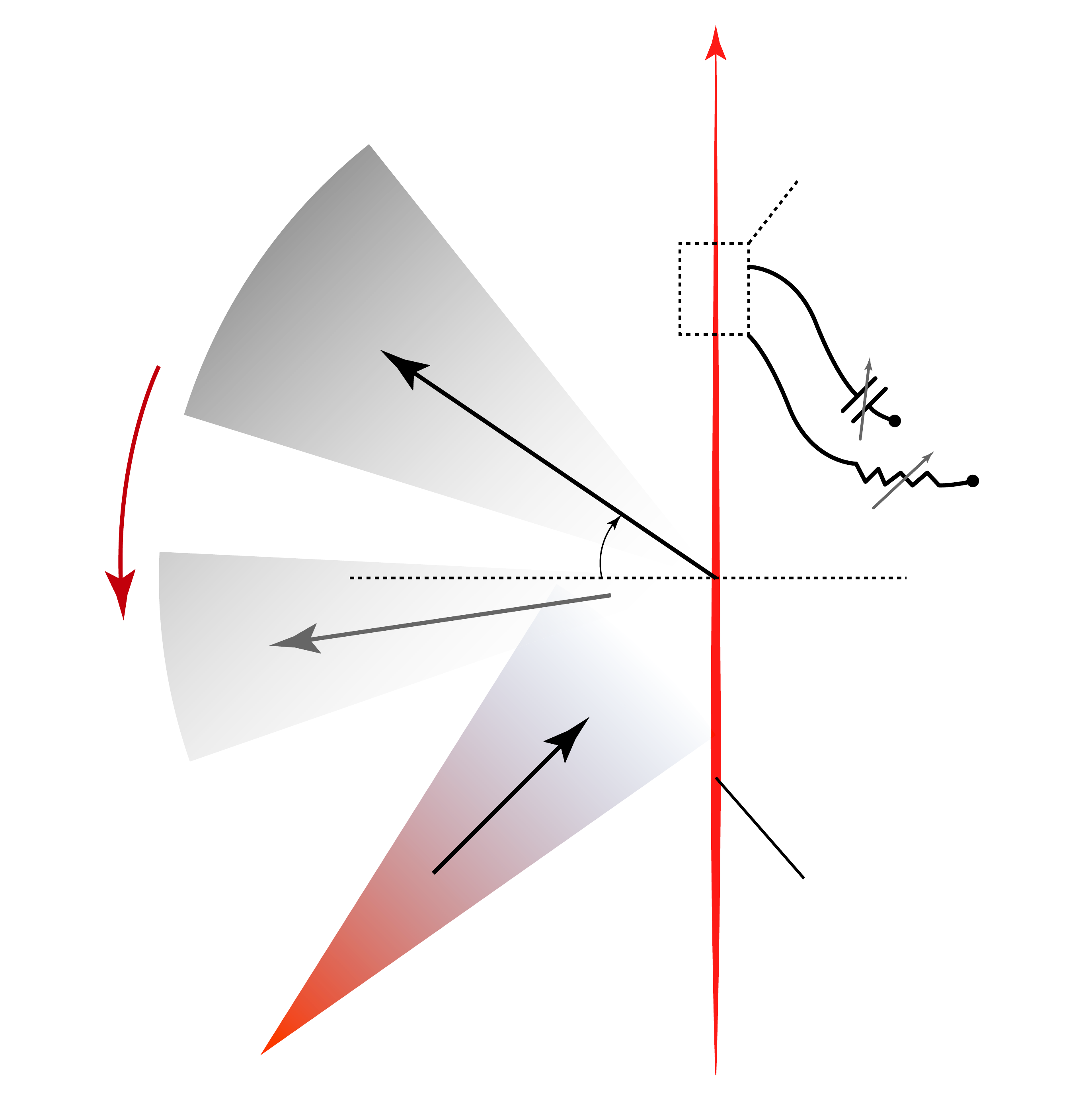}
			\put(85,14){\scriptsize \makebox(0,0){\textsc{\shortstack{Metasurface \\Reflector\\ $R(x)=|R(x)|e^{j\angle R(x)}$}}}}
			\put(20,21){\scriptsize \makebox(0,0){\textsc{\shortstack{Incident Wave \\ $\psi_0(\mathbf{r},\omega)$}}}}
			\put(13,85){\scriptsize \makebox(0,0){\shortstack{\textsc{Reflected Wave} \\ $\psi_\text{ref.}(\mathbf{r},V_{c1}, V_{r1})$}}}
			\put(0,37){\scriptsize \makebox(0,0){\shortstack{\textsc{Reflected Wave} \\ $\psi_\text{ref.}(\mathbf{r},V_{c2}, V_{r2})$}}}
			\put(64.5,100){\scriptsize \makebox(0,0){$x$}}
			\put(79,86){\scriptsize \makebox(0,0){Unit Cell}}
			\put(85,62){\scriptsize \makebox(0,0){$V_c^n$}}
			\put(92,57){\scriptsize \makebox(0,0){$V_r^n$}}
			\put(50,52){\scriptsize \makebox(0,0){$\theta_0$}}
	\end{overpic}
	\caption{Illustration of a reconfigurable metasurface reflector which transforms in incoming incident fields into desired scattered fields via active unit cells consisting of varactor diodes and PIN diodes for an independent phase and amplitude control, respectively.}
	\label{Fig:MS}
\end{figure}

Consider a reflection metasurface which is excited with an incident wave $\psi_0(\mathbf{r},\omega)$, as illustrated in Fig.~\ref{Fig:MS}. The metasurface is composed of an array of identical unit cells, where each unit cell is individually controlled with a tunable capacitor and a resistor with separate external voltage controls ($V_{c}, V_{r}$). The tunable lumped elements integrated into the unit cell controls the spatially varying complex reflectance, $\Gamma(x)=|\Gamma(x)|e^{j\angle \Gamma(x)}$, of the metasurface across the surface. By varying the voltage controls, we wish to reconfigure the reflected scattered fields, and in doing so, wish to devise a unit cell that is capable of providing a full range of reflection's amplitude $|\Gamma(f_0)|\in [0,~1]$ and phase $\angle \Gamma(f_0)\in[0,~2\pi]$ independently at a desired frequency of operation. 
\subsection{Need for Independent Magnitude/Phase Control}
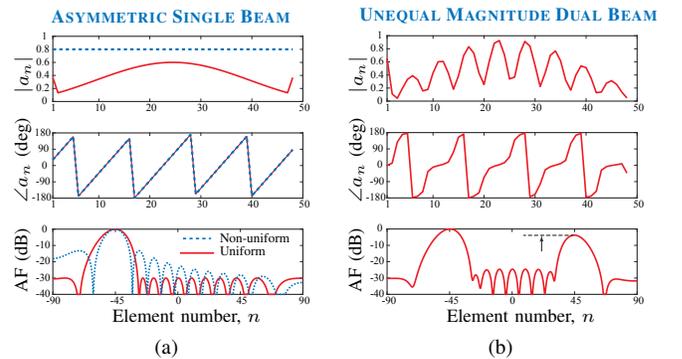
\begin{figure}[!b]
	\centering
	\subfigure[][]{
		\begin{overpic}[grid=false, width=0.46\columnwidth]{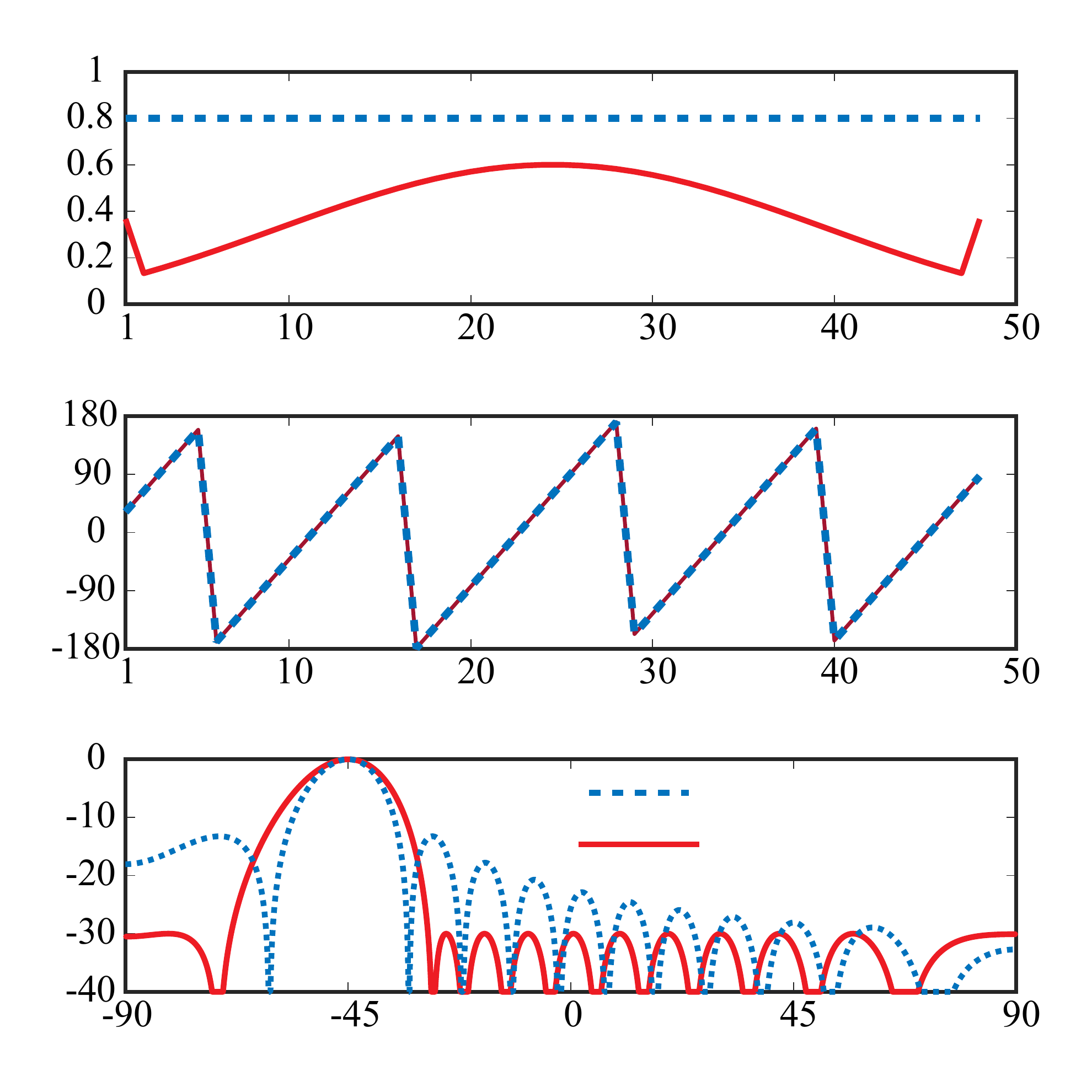}
			\put(-1,75){\rotatebox{90}{\scriptsize $|a_n|$}}
			\put(-1,39){\rotatebox{90}{\scriptsize $\angle a_n$ (deg)}}
			\put(31,0){\scriptsize Element number, $n$}
			\put(-1,10){\rotatebox{90}{\scriptsize AF (dB)}}
			\put(50,100){\makebox(0,0){\color{matlabblue}\scriptsize \textbf{\textsc{\shortstack{Asymmetric Single Beam}}}}}
			\put(66,22){\tiny Uniform}
			\put(66,26){\tiny Non-uniform}
		\end{overpic}
		\label{two_beam_gen_exp_1}}
	\subfigure[][]{
		\begin{overpic}[grid=false, width=0.46\columnwidth]{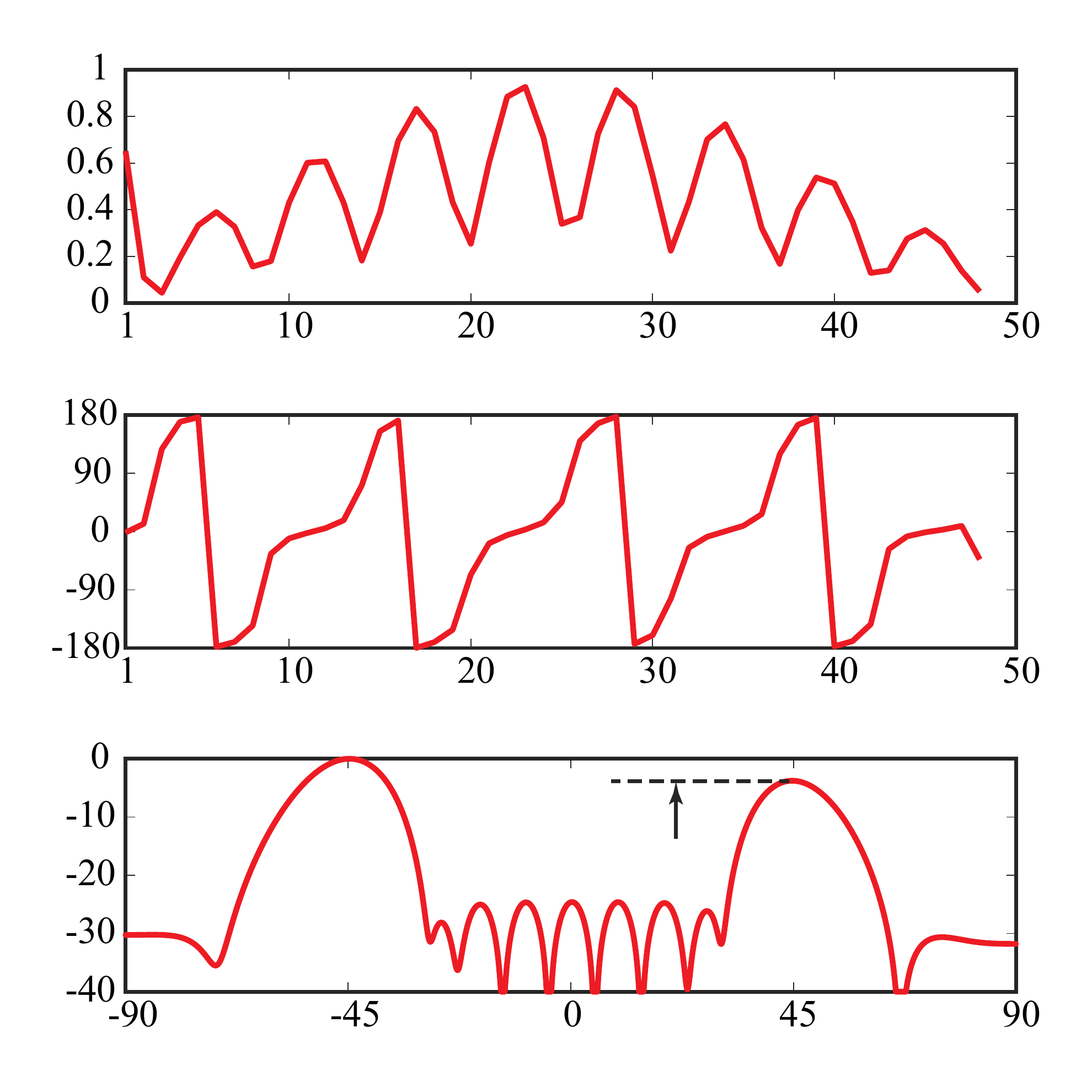}
			\put(-1,75){\rotatebox{90}{\scriptsize $|a_n|$}}
			\put(-1,39){\rotatebox{90}{\scriptsize $\angle a_n$ (deg)}}
			\put(31,0){\scriptsize Element number, $n$}
			\put(-1,10){\rotatebox{90}{\scriptsize AF (dB)}}
			\put(50,100){\makebox(0,0){\color{matlabblue}\scriptsize\textbf{ \textsc{\shortstack{Unequal Magnitude Dual Beam}}}}}
		\end{overpic}
		\label{two_beam_gen_exp_2}}
	\caption{Illustrations on the importance of controlling the array's elements magnitudes showing the Array Factors (AFs) of a uniform spaced linear array with uniform and non-uniform amplitude distributions with a specific phase progression. (a) The non-uniform and uniform AFs for a single tilted beam at an angle of 45$^{\circ}$ and (b) A multi-beam array tilted at angles of +/-45$^{\circ}$ with a different reflection gain for each beam.}
	\label{two_beam_gen_exp_1_n_2}
\end{figure}
To illustrate the requirement and importance of controlling reflection magnitude and phase, consider that the metasurface is excited with a normally incident uniform plane-wave, for simplicity. The reflected scattered fields in the far-field of the surface due to spatially varying complex reflectance of the surface, can be constructed using the standard array factor of antenna theory (linear polarization with no rotation is assumed). More specifically, considering the metasurface as an $N-$element linear array of unit cell size $d$, the overall reflection pattern of the surface (considering the surface as a distribution of point sources) may simply be modeled as
	\begin{equation}
	AF = \sum_{n=1}^{N} a_n e^{j(n-1)(kd\cos{\theta} +\beta)}
	\label{Eq:AF}
	\end{equation}
\noindent where $a_n$'s are the complex excitation coefficients for each element, $\beta$ is the progression phase between the array's elements (which is zero for normally incident plane-wave, assumed here), $k$ is the wavenumber, $\theta$ is the angle between the axis of the array along the $x-$axis and the radial vector from the origin to the observation point.

Let us take an example of a beam-steered metasurface which reflects the incident beam at an angle of $\theta_0$. As well-known from both antenna theory and metasurface analysis, a linear phase gradient across the metasurface will provide such a beam-tilt in the far-field. Fig.~\ref{two_beam_gen_exp_1_n_2}(a) shows an example, where the beam is reflected off $\theta_0 = -45\degree$ using a metasurface of finite size, where a linear phase tilt and a uniform magnitude profile is imposed across the surface. As expected from a uniform magnitude surface, the main radiation beam is accompanied by side-lobes with peak-values at $-13$~dB, typical of uniform apertures. If now instead, a spatially varying non-uniform magnitude profile is also imposed in addition to the phase, the side-lobes in the reflection field can be engineered and greatly reduced\footnote{Several amplitude tapering schemes exist in standard antenna theory textbooks on array synthesis. Chebyshev arrays is used here as an example.}, as shown in Fig.~\ref{two_beam_gen_exp_1_n_2}(a) maintained at $-30$~dB, for instance. Another case that exemplifies the importance of a simultaneous magnitude and phase control is that of designing two asymmetrically located beams. This example is shown in Fig.~\ref{two_beam_gen_exp_1_n_2}(b), where the incident wave is split into two beams, with unequal magnitudes. It is clear, that such a reflection pattern cannot be achieved using either phase or magnitude control only.
\begin{figure}[hbtp]
	\centering
	\subfigure[][]{
	\begin{overpic}[grid=false, width=0.85\columnwidth]{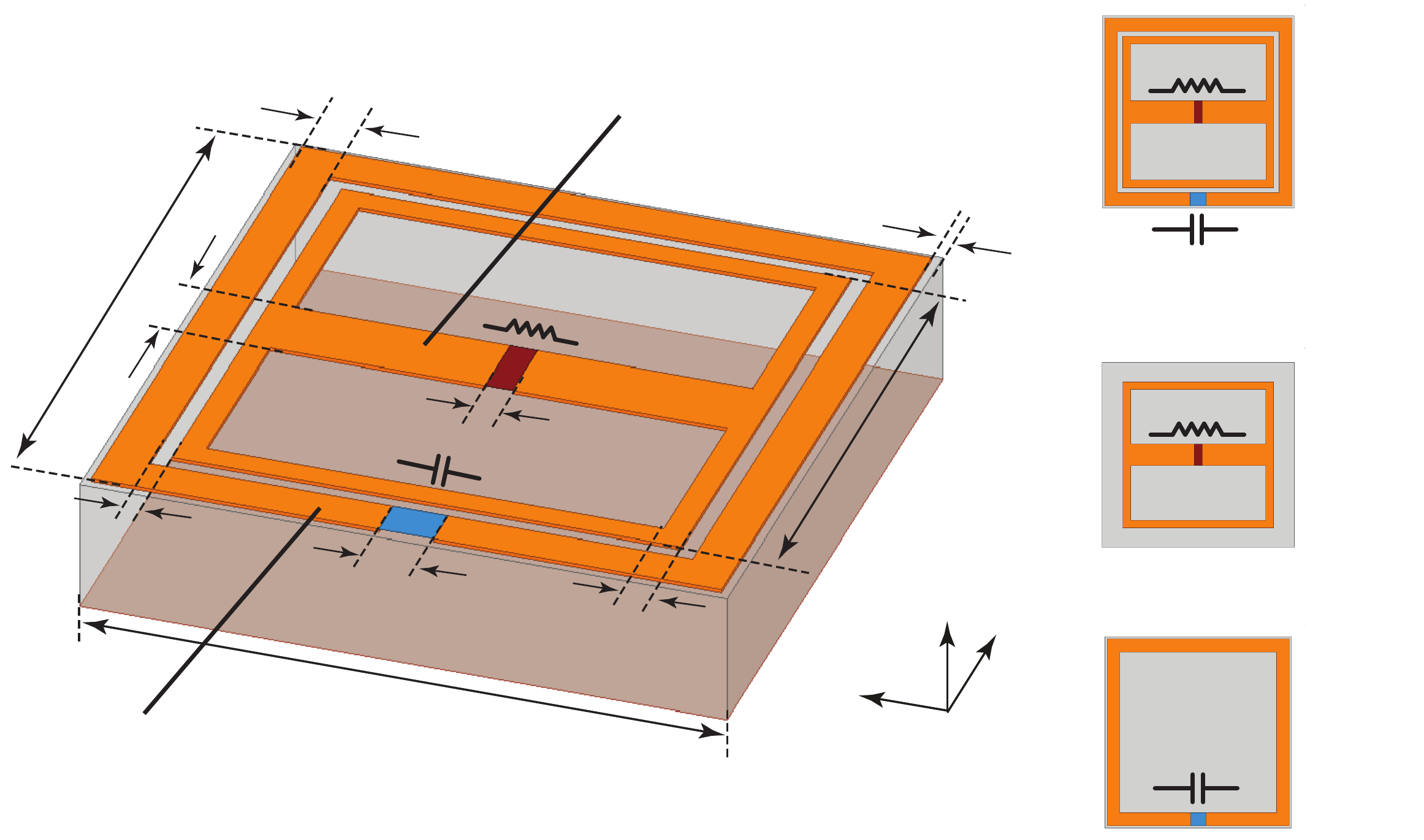}
		\put(99,8){\makebox(0,0){\color{matlabblue}\scriptsize \shortstack{Split-Ring \\Only}}}
		\put(100,27){\makebox(0,0){\color{matlabblue}\scriptsize \shortstack{Dipole-Ring \\Only}}}
		\put(98.5, 52){\makebox(0,0){\color{amber}\scriptsize \textbf{\shortstack{Coupled \\Resonator}}}}
		\put(84,38){\makebox(0,0){\large$=$}}
		\put(84,17){\makebox(0,0){\large$+$}}
		
		\put(6,40){\scriptsize $a$}
		\put(24,52){\scriptsize $b$}
		\put(61,25){\scriptsize $c$}
		\put(42,13){\scriptsize $d$}
		\put(10,37){\scriptsize $e$}
		\put(7,19){\scriptsize $f$}
		\put(32,27){\scriptsize $g$}
		\put(26,16){\scriptsize $l$}
		\put(67,45){\scriptsize $s$}
		\put(24,8){\scriptsize $\Lambda$}
		
		\put(57,10){\scriptsize $y$}
		\put(65.5,16){\scriptsize $z$}
		\put(70,15){\scriptsize $x$}
		\put(10,2){\makebox(0,0){\color{matlabblue}\scriptsize \shortstack{ Split-Ring Resonator \\(SRR) with \\ \textbf{Tunable Capacitor}}}}
		\put(56,56){\makebox(0,0){\color{matlabblue}\scriptsize \shortstack{ Dipole-Ring Resonator \\ (DRR) with \\ \textbf{Tunable Resistor}}}}
	\end{overpic}
	\label{ms_prespective_2}}
	
		\subfigure[][]{
		\begin{overpic}[grid=false, width=1\columnwidth]{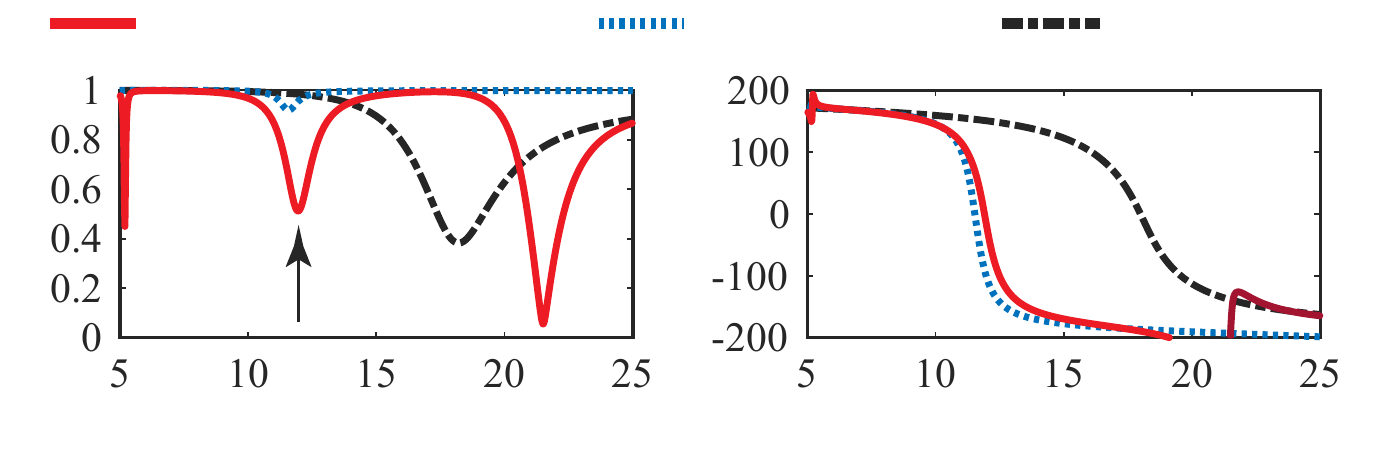}
			\put(18,1){\scriptsize Frequency (GHz)}
			\put(0,9){\rotatebox{90}{\scriptsize Magnitude, $|\Gamma|$}}
			\put(68,1){\scriptsize Frequency (GHz)}
			\put(48,8){\rotatebox{90}{\scriptsize Phase, $\angle \Gamma$ (deg)}}	
			
			\put(11, 30){\scriptsize Coupled Resonator}
			\put(50, 30){\scriptsize SRR}
			\put(81,30){\scriptsize DRR}
		\end{overpic}
		\label{hfss_DRR_vs_SRR_vs_DRR_SRR_RC}}
	
		\subfigure[][]{
		\begin{overpic}[grid=false, width=1\columnwidth]{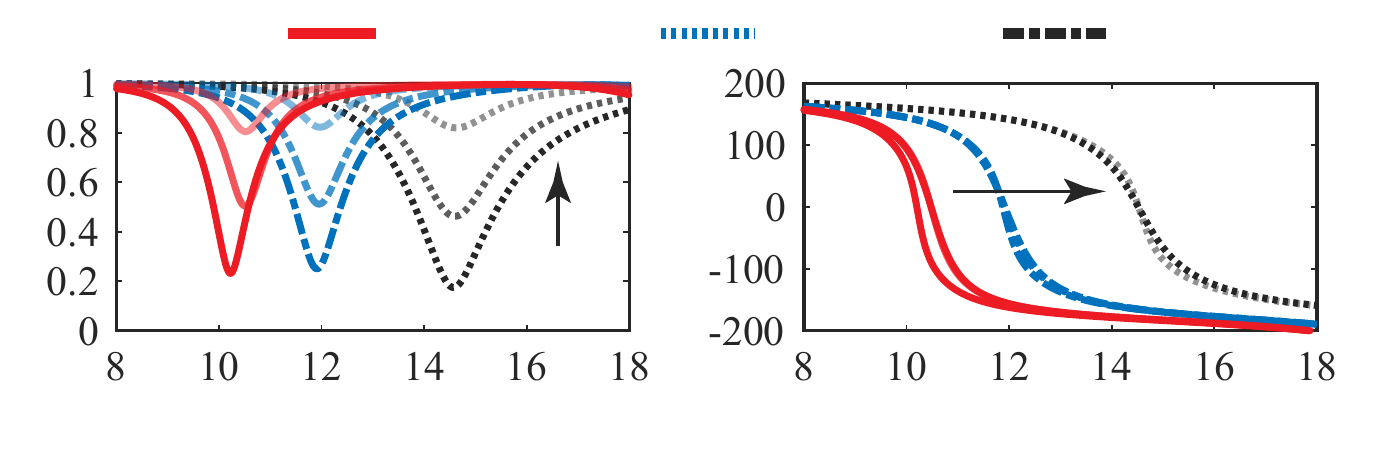}
			\put(18,2){\scriptsize Frequency (GHz)}
			\put(0,10){\rotatebox{90}{\scriptsize Magnitude, $|\Gamma|$}}
			\put(68,2){\scriptsize Frequency (GHz)}
			\put(48,8){\rotatebox{90}{\scriptsize Phase, $\angle \Gamma$ (deg)}}
			
			\put(28,29.5){\fontsize{6}{2}$\text{C = 0.25 pF}$}
			\put(55,29.5){\fontsize{6}{2}$\text{C = 0.1 pF}$}
			\put(81,29.5){\fontsize{6}{2}$\text{C = 0.025 pF}$}
			\put(83,18){\tiny \color{amber}\shortstack{Decreasing\\ Capacitance}}
			\put(35.5,10){\tiny \color{amber} \shortstack{Increasing\\ Resistance}}
		\end{overpic}
		\label{hfss_DRR_SRR_RC}}
	\caption{Proposed metasurface unit cell architecture based on coupled Split Ring Resonator (SRR) and Dipole Ring Resonator (DRR), loaded with tunable capacitor and resistor respectively. (b) FEM-Simulated magnitude and phase of three configuration ($R = 25~\Omega$ and $C$ = 0.1~pF). (c) Simulated magnitude and phase of the coupled-resonator unit cell with three different resistances ($R = 5, 25, 50~\Omega$) and three different capacitances (C = 0.025, 0.1, 0.25~pF). All simulations performed in FEM-HFSS, and the various dimensions are provided in Tab.~I.}
\end{figure}
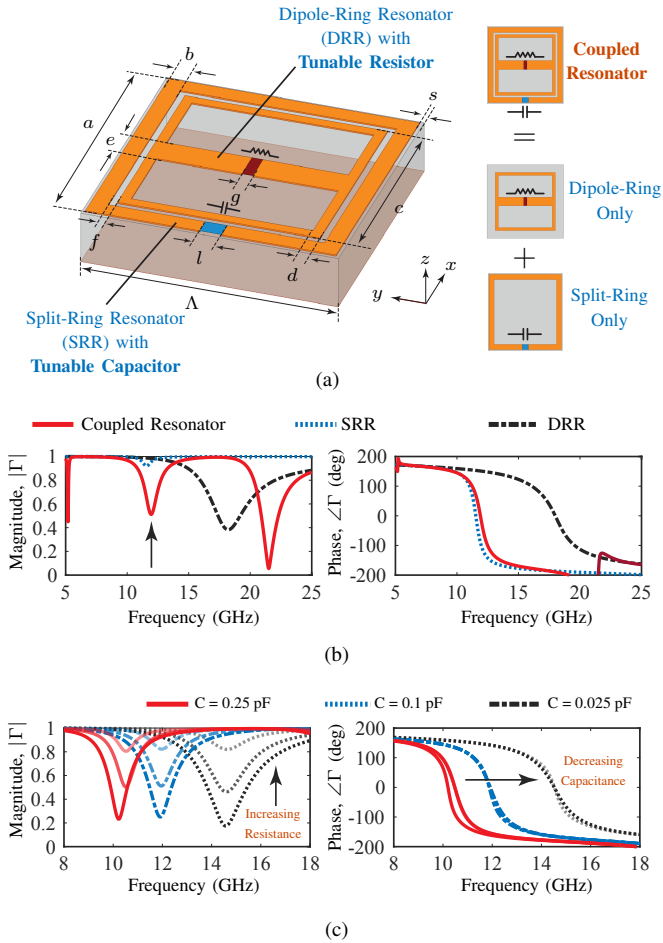%
\subsection{Proposed Coupled-Resonator Structure}

A conventional reflection unit cell providing real-time phase control consists of a lumped tuning element, typically a varactor diode for instance at Radio Frequencies (RF), which is integrated inside a sub-wavelength resonator. Two simple geometries that are commonly used are Split Ring Resonator (SRR) and Dipole Ring Resonator (DRR), as shown in Figure~\ref{ms_prespective_2}. As their resonance frequency is tuned via lumped capacitance, the reflection phase at a desired fixed frequency is changed along with an uncontrolled magnitude variation due to dissipation losses of the materials. For a controlled magnitude variation, a second dissipative element must be added that controls the resonator losses, which should ideally operate independently of the capacitance value.

Figure~\ref{ms_prespective_2} further shows the proposed unit cell formed as a combination of an SRR and DRR, incorporating these two lumped elements for achieving independent amplitude and phase control. The architecture specifically consists of an SRR loaded with a varactor acting as a \emph{tunable capacitor} which dominantly controls its resonant frequency. A second DRR is inserted inside the SRR loaded with a PIN diode, acting as a \emph{tunable resistor} which controls the reflection magnitude of the cell\footnote{While SRR and DRR are used here for convenience, the unit cell may be adapted with more sophisticated resonator geometries for greater wave control such as manipulating polarization, for instance.}. Since the two resonators are \emph{strongly coupled}, the unit cell is operated at one of its \emph{coupled resonances}, so that its reflection magnitude and phase is controlled by the two lumped tuning elements together. In all cases, the unit cell period $\Lambda \ll \lambda_0$ to maintain sub-wavelength characteristics for good spatial discretization of magnitude and phase when configured to form a non-uniform metasurface.

A typical unit cell response of the proposed unit cell is shown in Fig.~\ref{hfss_DRR_vs_SRR_vs_DRR_SRR_RC}, along with those of isolated SRR and DRR structures, loaded with capacitance $C$ and resistance $R$, respectively. Owing to the larger resonant length, the SRR features a lower resonant frequency (dashed blue curve) compared to that of DRR (dashed black curve) when working in isolation, and thus have different frequency of resonances. When they are superimposed, they are strongly coupled electromagnetically, and the resonant frequencies of the combined unit cell are strongly perturbed, as seen in Fig.~\ref{hfss_DRR_vs_SRR_vs_DRR_SRR_RC}. 
\begin{figure}[!t]
	\centering
		\subfigure[][]{
	\begin{overpic}[grid=false, width=1\columnwidth]{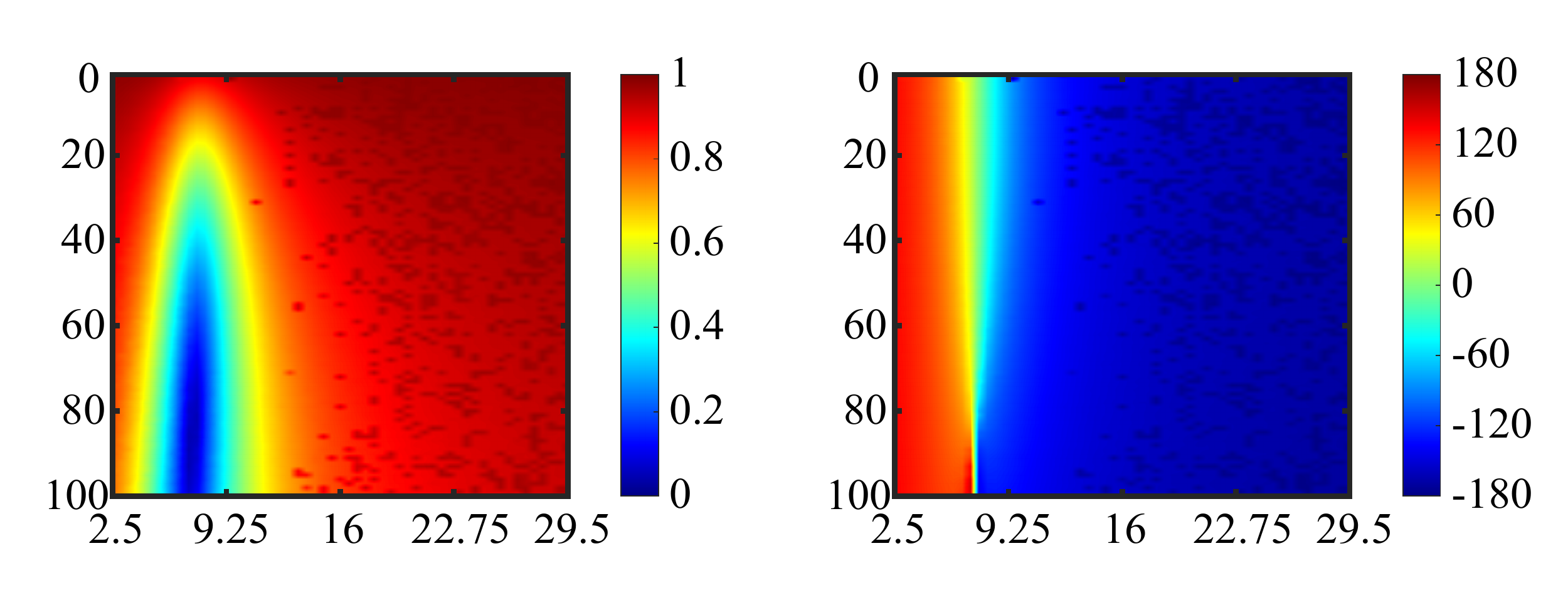}
		\put(7,0){\scriptsize Capacitance $\times10^{-2}$(pF)}
		\put(0,12){\rotatebox{90}{\scriptsize Resistance~($\Omega$)}}
		\put(57,0){\scriptsize Capacitance $\times10^{-2}$(pF)}
		\put(50,12){\rotatebox{90}{\scriptsize Resistance~($\Omega$)}}
		\put(68,12){\scriptsize \color{white} \shortstack{(Unit Cell\\FEM-HFSS)}}
		\put(8.5,36){\scriptsize $|\Gamma|~(f=12.5~\text{GHz})$}
		\put(57.5,36){\scriptsize $\angle \Gamma~(f=12.5~\text{GHz})$}
		\put(-3,-27){\rotatebox{90}{\makebox(0,0){\footnotesize \color{amber} \textsc{\textbf{Single Resonator (DRR)}}}}}
	\end{overpic}
		\label{Amp_Phs_RC_two_resonators}}\hfill
	\subfigure[][]{
	\begin{overpic}[grid=false, width=1\columnwidth]{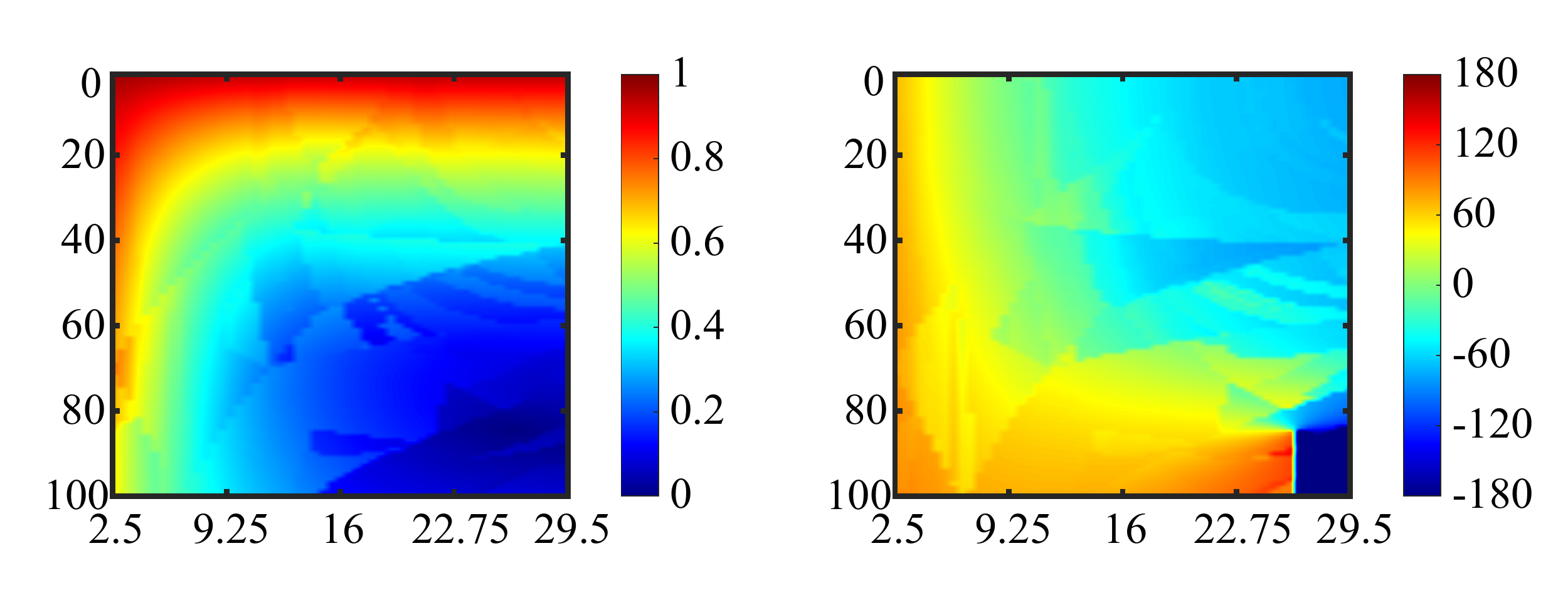}
		\put(7,0){\scriptsize Capacitance $\times10^{-2}$(pF)}
		\put(0,12){\rotatebox{90}{\scriptsize Resistance~($\Omega$)}}
		\put(57,0){\scriptsize Capacitance $\times10^{-2}$(pF)}
		\put(50,12){\rotatebox{90}{\scriptsize Resistance~($\Omega$)}}
		\put(8.5,36){\scriptsize $|\Gamma|~(f=18.5~\text{GHz})$}
		\put(57.5,36){\scriptsize $\angle \Gamma~(f=18.5~\text{GHz})$}
		\put(-3,67){\rotatebox{90}{\makebox(0,0){\footnotesize \color{cyan} \textsc{\textbf{Coupled Resonator}}}}}
		\put(68,25){\scriptsize\shortstack{(Unit Cell\\FEM-HFSS)}}
	\end{overpic}
		\label{Amp_Phs_RC_singel_ring_dipole_series}}
	\caption{FEM-HFSS simulated amplitude-phase distribution at a fixed frequency as a function of lumped element values. (a) Proposed coupler resonator unit cell. (b) Dipole Ring Resonator unit cell where the resistance and capacitance are configured in series in its gap. All simulations performed in FEM-HFSS, and the various dimensions are provided in Tab.~I.}
	\label{Amp_Phs_RC_singel_ring_dipole_series_&_Amp_Phs_RC_two_resonators}
\end{figure}

\begin{figure}[!h]
	\centering
	\subfigure[][]{
	\begin{overpic}[grid=false, width=0.45\columnwidth]{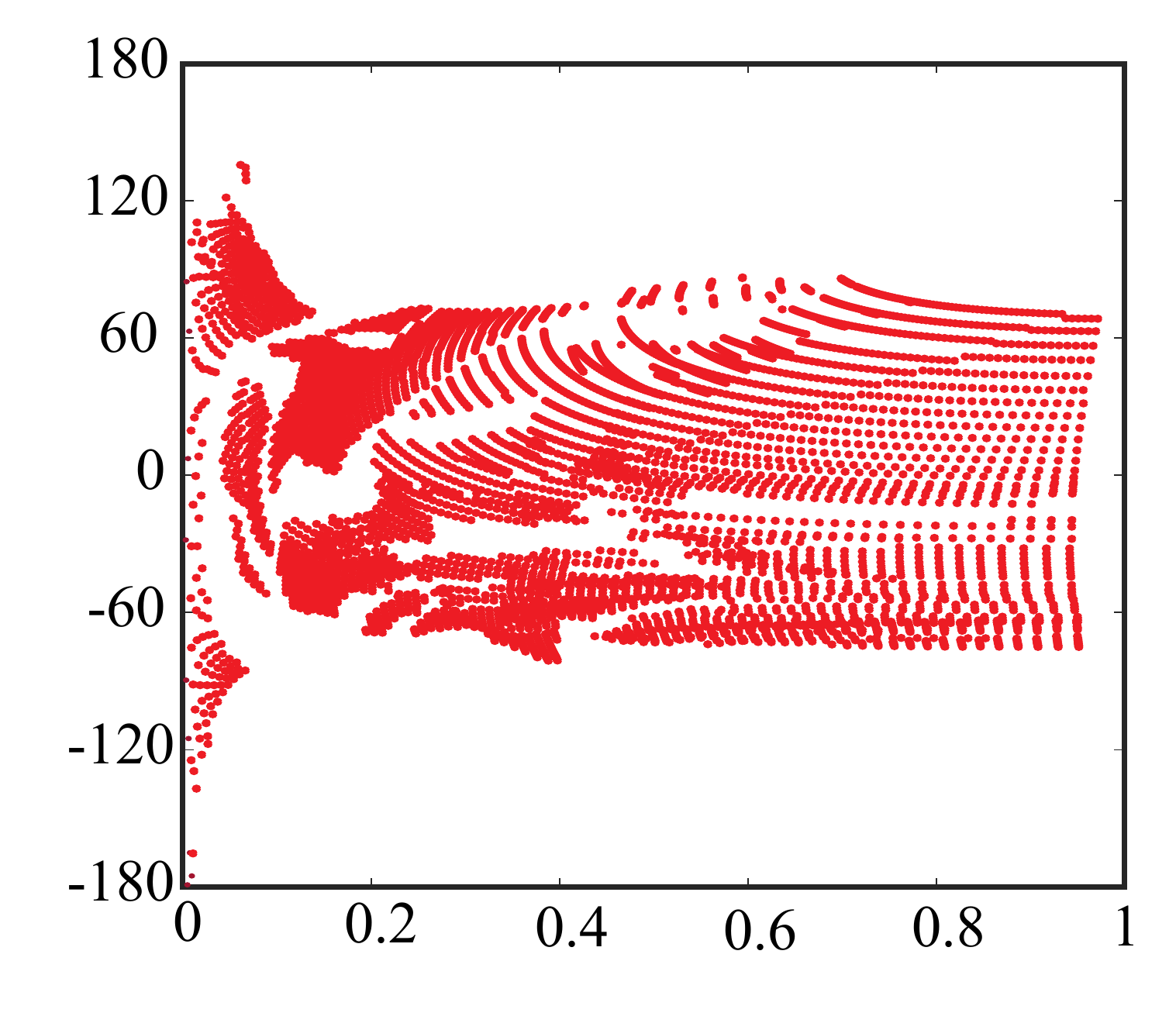}
		\put(40,0){\scriptsize Magnitude, $|\Gamma|$}
		\put(0,30){\rotatebox{90}{\scriptsize phase, $\angle \Gamma$ (deg)}}
		\put(55,90){\makebox(0,0){\footnotesize \color{amber} \textsc{\textbf{Single Resonator (DRR)}}}}

	\end{overpic}
		\label{Amp_Phs_RC_dipole_ring_series_unique}}
	\subfigure[][]{
	\begin{overpic}[grid=false, width=0.45\columnwidth]{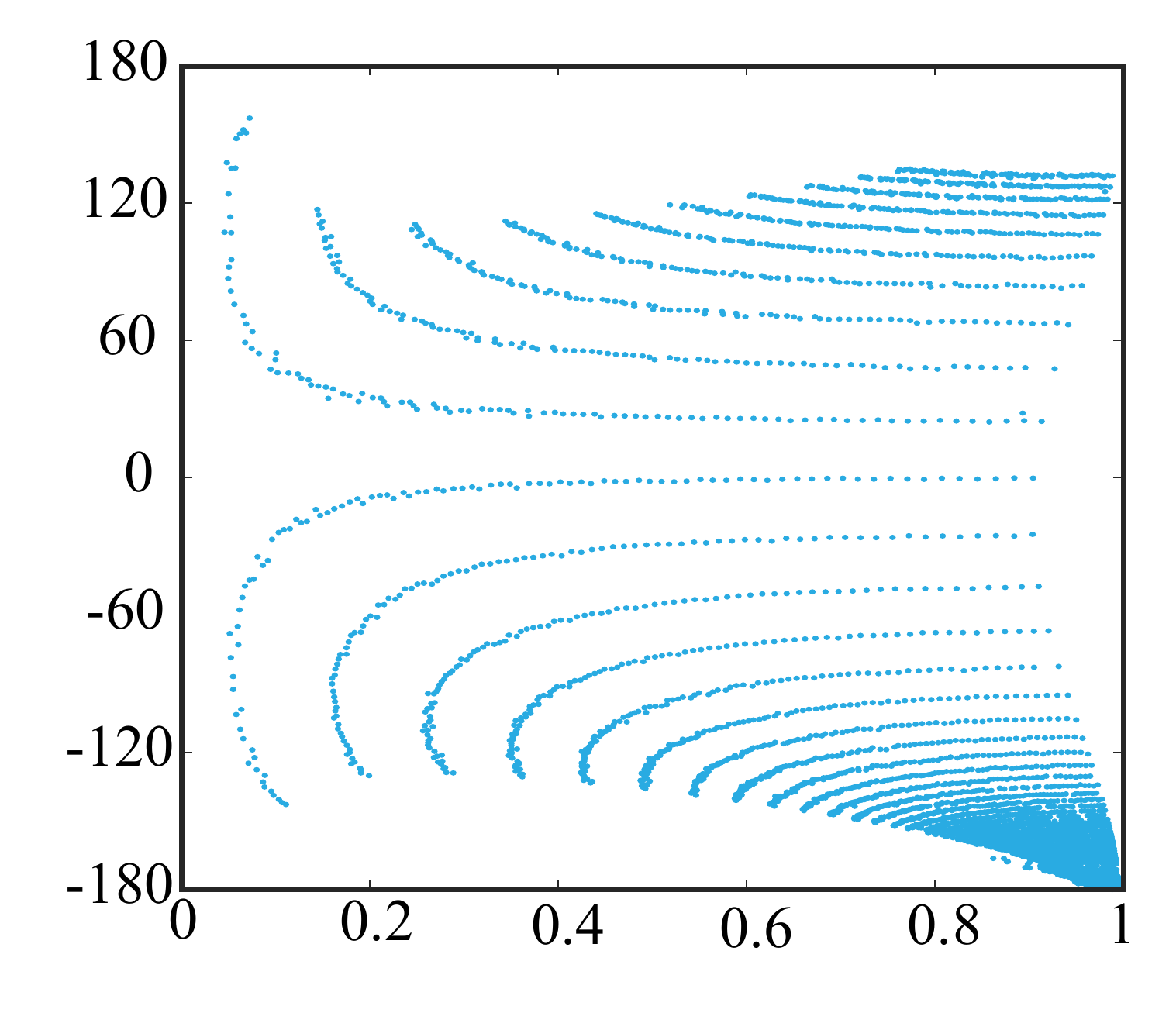}
		\put(40,0){\scriptsize Magnitude, $|\Gamma|$}
		\put(0,30){\rotatebox{90}{\scriptsize phase, $\angle \Gamma$ (deg)}}
		\put(55,90.5){\makebox(0,0){\footnotesize \color{cyan} \textsc{\textbf{Proposed Coupled Resonator}}}}
	\end{overpic}
		\label{Amp_Phs_RC_two_resonators_unique}}
	\caption{Amplitude-phase coverage plots for the unit cells of Fig.~\ref{Amp_Phs_RC_singel_ring_dipole_series_&_Amp_Phs_RC_two_resonators} showing achievable values using the capacitance values ranging between $0.025 - 0.295$~pF and resistance values ranging between $1 - 100$~$\Omega$.(a) A single dipole-ring resonator of Fig.~\ref{Amp_Phs_RC_singel_ring_dipole_series}. (b) Proposed coupled-resonator of Fig.~\ref{Amp_Phs_RC_two_resonators}.} 
	\label{Amp_Phs_RC_dipole_ring_series_parallel_&_Amp_Phs_RC_two_resonators_unique}
\end{figure}
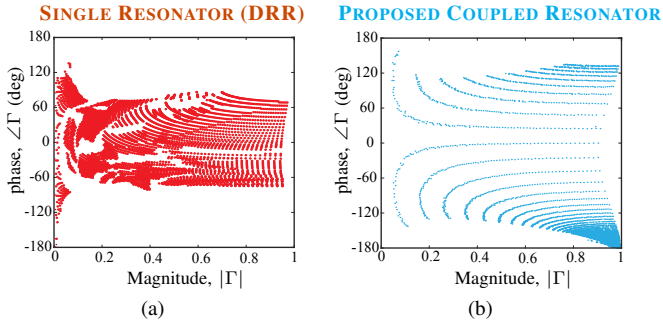
Now, we can choose to operate at one of these resonant frequencies knowing that the resonant frequency $f_0$ and the Q-factor depend on both the capacitance and resistance values. This is illustrated in Fig.~\ref{hfss_DRR_SRR_RC} using one of the resonant peaks of the coupled system. For a fixed capacitance value $C$, when the resistance of the DRR is changed, the Q-factor ($=f_0/\Delta f$) changes with minimal effect on $f_0$. Larger $R$ is seen to correspond to lower reflection from the surface, while maintaining the resonance frequency and thus the reflection phase. As the capacitance is decreased, the resonant frequency $f_0$ is increased as expected, while nearly maintaining its reflection magnitude level. Thus $C$ and $R$ represent two independent controls on the reflection phase and magnitude with minimal interdependence. 

In practical cases, the operation frequency $f_0$ is typically fixed and thus it is preferable to obtain the amplitude-phase map achievable through a given unit cell configuration. Fig.~\ref{Amp_Phs_RC_two_resonators} shows a 2D map of the reflection magnitude and phase for varying $R$ and $C$ at a fixed $f_0$. The reflection phase shows only a slight variation as the resistance value is changed for a fixed capacitance, while large variation is seen in the magnitude response. On the other hand, the reflection phase is gradually decreased as Capacitance increases, except near the regions where the surface is fully absorptive, i.e. $|\Gamma|\approx 0$. If a specific reflection amplitude-phase pair $\{|\Gamma|, \angle \Gamma\}$ is sought, the phase is first fixed to $\angle \Gamma$ by choosing an appropriate capacitance $C_0$, followed by resistance $R_0$ tuning to achieve the desired reflection magnitude of $|\Gamma|$.

At this point, one may wonder if a single SRR or DRR resonator with both $R$ and $C$ may also provide an independent amplitude and phase control? To investigate this, Fig.~\ref{Amp_Phs_RC_two_resonators_unique} shows the 2D reflection amplitude and phase maps for a DRR unit cell where a series $R-C$ is integrated on the gap. In this case, large phase tuning is observed at a higher frequency for the same lumped element ranges as that of Fig.~\ref{Amp_Phs_RC_two_resonators} (i.e. 18.5~GHz vs 12.5~GHz, around the isolated resonance of the DRR). No clear independent control is visible, although some combinations of $R-C$ may be found which can provide desired $\{|\Gamma|,\angle \Gamma\}$ combinations. This amplitude-phase coverage range, may more clearly be seen in Fig.~\ref{Amp_Phs_RC_dipole_ring_series_parallel_&_Amp_Phs_RC_two_resonators_unique}, where each pixel represents amplitude-phase combinations which are possible to achieve using a certain $\{R,C\}$ combination. While a single resonator configuration shows a dense distribution with a large magnitude variation, the phase range is restricted, in addition to a lack of a clear mapping between $\{R,C\}$ and $\{|\Gamma|,\angle \Gamma\}$. On the other hand, the combined proposed unit cell features a well-defined distribution with significantly larger coverage across both phase and magnitude.

It should further be noted that even though some level of amplitude-phase control is possible using a single resonator-based unit cell, it is not particularly suited for practical implementation. Both the varactor and the PIN diodes as practical means of controlling capacitance and resistance require separate voltage control lines for reverse-biasing and forward-biasing, respectively, which is not convenient in a single resonator configuration. On the other hand, for the proposed coupled-resonator cells, since the two lumped elements are physically disconnected and located on different resonators, they can be individually biased using standard biasing networks with no practical difficulty. Therefore, the proposed coupled-resonator architecture is clearly superior to a single resonator-based one from both electrical as well as practical implementation point of view.
\begin{figure}[!b]
	\centering
	\subfigure[][]{
		\begin{overpic}[grid=false, scale=0.26]{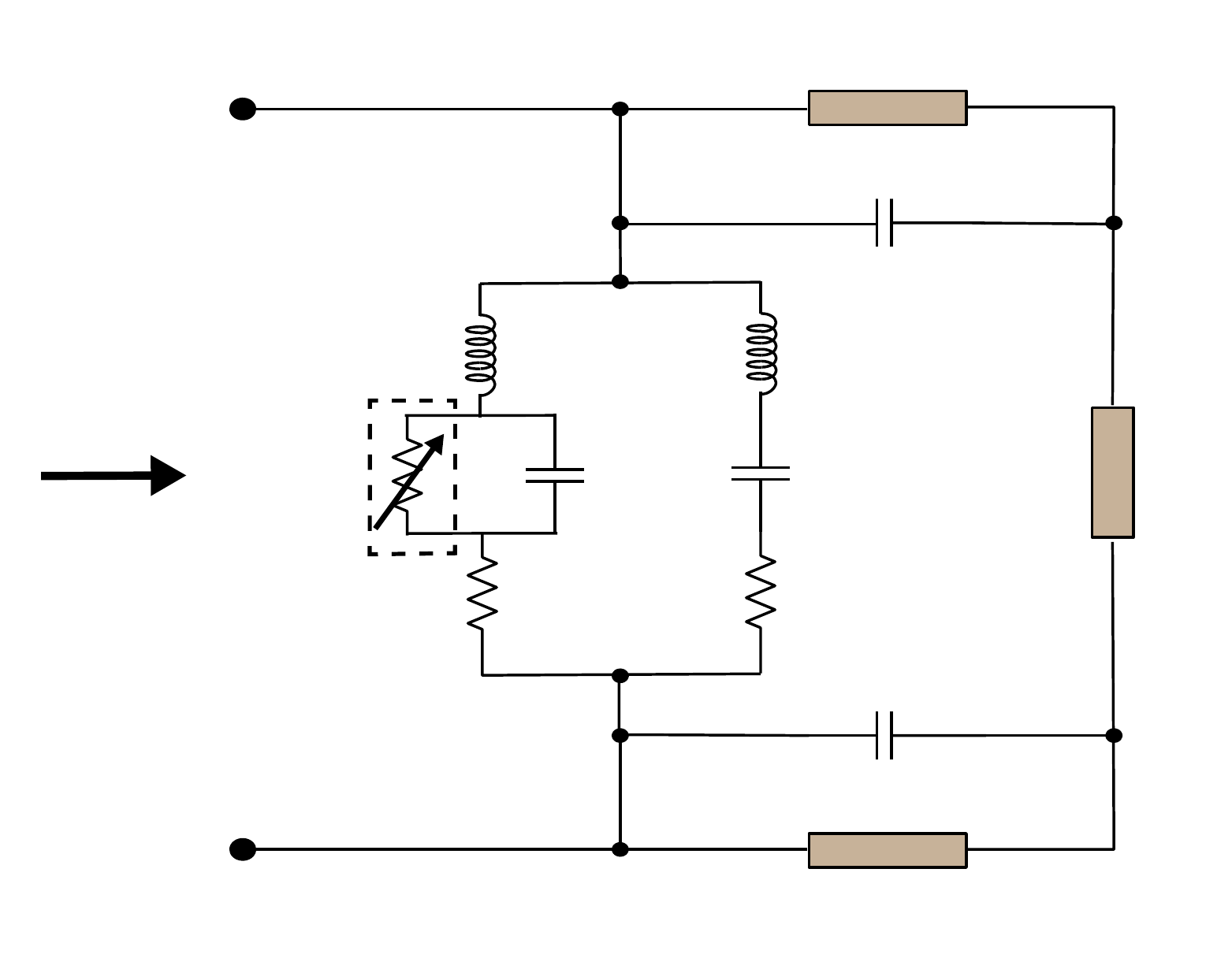}
		  \put(32,47){\scriptsize $l_1$}
		  \put(65,47){\scriptsize $l_2$}
		  \put(48,38){\scriptsize $c_1$}
		  \put(65,38){\scriptsize $c_2$}
		  \put(32,28){\scriptsize $r_1$}
		  \put(65,28){\scriptsize $r_2$}
		  \put(69,54){\scriptsize $c_d$}
		  \put(69,22){\scriptsize $c_d$}
		  \put(24,37){\color{red}\fontsize{6}{4}$R$}
		  \put(0,43){\fontsize{6}{4}\shortstack{$\Gamma(f_0)$ \\ $Z_0$}}
		  \put(67,72){\fontsize{6}{4}$Z, \theta$}
		  \put(67,2){\fontsize{6}{4}$Z, \theta$}
		  \put(94,48){\rotatebox{-90}{\fontsize{6}{4}$Z_\text{short}$}}
			\put(50,82){\makebox(0,0){\scriptsize \color{amber}\textsc{\textbf{Dipole Ring Resonator}}}}
			\put(157,82){\makebox(0,0){\scriptsize \color{amber}\textsc{\textbf{Split Ring Resonator}}}}
		\end{overpic}
		\label{dipole_ring_ckt_model}}
	\subfigure[][]{
		\begin{overpic}[grid=false, scale=0.25]{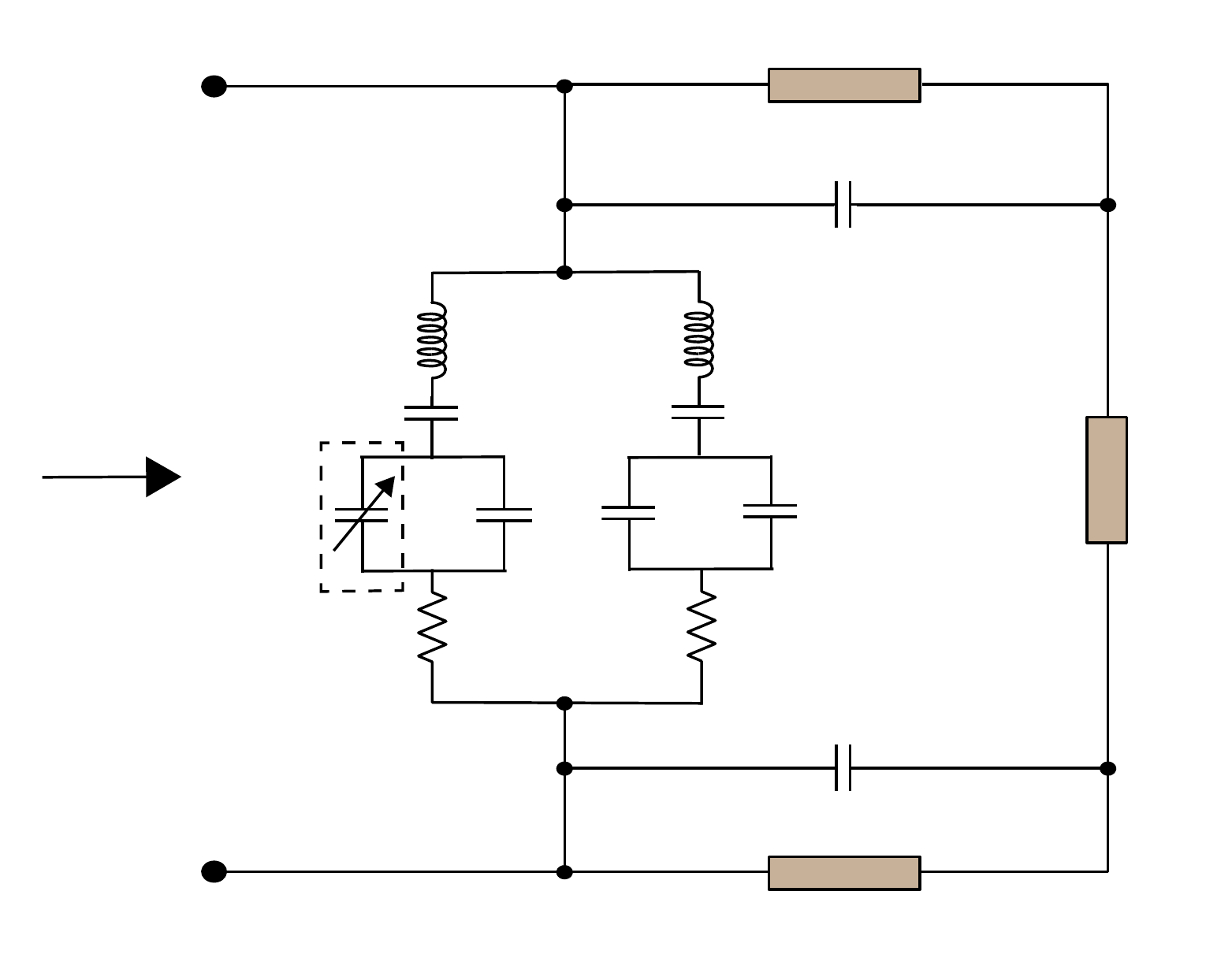}
			\put(29,47){\scriptsize $l_1$}
			\put(60,47){\scriptsize $l_2$}
			\put(38,43){\scriptsize $c_1$}
			\put(48,43){\scriptsize $c_2$}
			\put(43.5,34){\scriptsize $c_3$}
			\put(53.5,34){\scriptsize $c_4$}
			\put(27,25){\scriptsize $r_1$}
			\put(59,25){\scriptsize $r_2$}
			\put(66,55){\scriptsize $c_s$}
			\put(66,19){\scriptsize $c_s$}
			\put(20,34){\color{red}\fontsize{6}{4}$C$}
			\put(65,34){\fontsize{6}{4}$k$\color{red}$C$}
			\put(0,42){\fontsize{6}{4}\shortstack{$\Gamma(f_0)$ \\ $Z_0$}}
			\put(63,74){\fontsize{6}{4}$Z, \theta$}
			\put(63,1){\fontsize{6}{4}$Z, \theta$}
			\put(94,48){\rotatebox{-90}{\fontsize{6}{4}$Z_\text{short}$}}
		\end{overpic}
		\label{ring_ckt_model}}
	\subfigure[][]{
	\begin{overpic}[grid=false, width=\columnwidth]{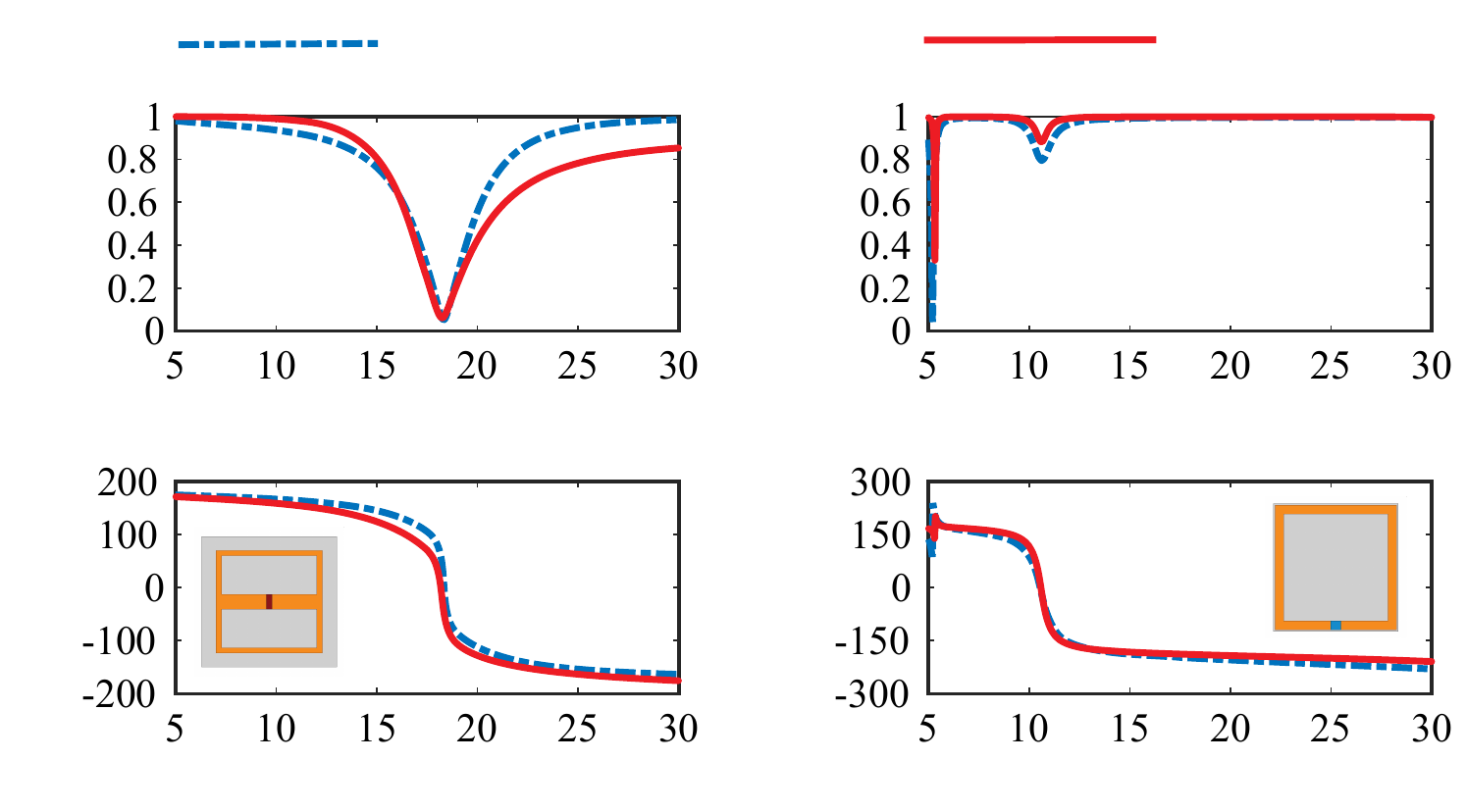}
		\put(20,2){\scriptsize Frequency (GHz)}
		\put(2,6){\rotatebox{90}{\scriptsize Phase, $\angle \Gamma$ (deg)}}		
		\put(71,2){\scriptsize Frequency (GHz)}
		\put(54,6){\rotatebox{90}{\scriptsize Phase, $\angle \Gamma$ (deg)}}		
		\put(20,27){\scriptsize Frequency (GHz)}
		\put(2,31){\rotatebox{90}{\scriptsize Magnitude, $|\Gamma|$}}		
		\put(71,27){\scriptsize Frequency (GHz)}
		\put(54,31){\rotatebox{90}{\scriptsize Magnitude, $|\Gamma|$}}		
		\put(27,50){\scriptsize \shortstack{Equivalent \\Circuit Model}}
		\put(80,50){\scriptsize \shortstack{Unit Cell \\FEM-HFSS}}
	\end{overpic}
	\label{hfss_vs_ckt_dipole_n_ring}}
	\caption{Developed equivalent circuit models of (a) the DRR and (b) the SRR with tunable resistance and capacitance, respectively. (c) Comparison of FEM-HFSS simulation with that of circuit model for the DRR (left) and SRR (right). The various circuit's parameters are summarized in Tab.~\ref{full_wave_ckt_spec}.}
	\label{hfss_vs_ckt_mod_dipole_n_ring}
\end{figure}
\begin{figure}[!t]
	\centering
	\subfigure[][]{
		\begin{overpic}[grid=false, width=0.9\columnwidth]{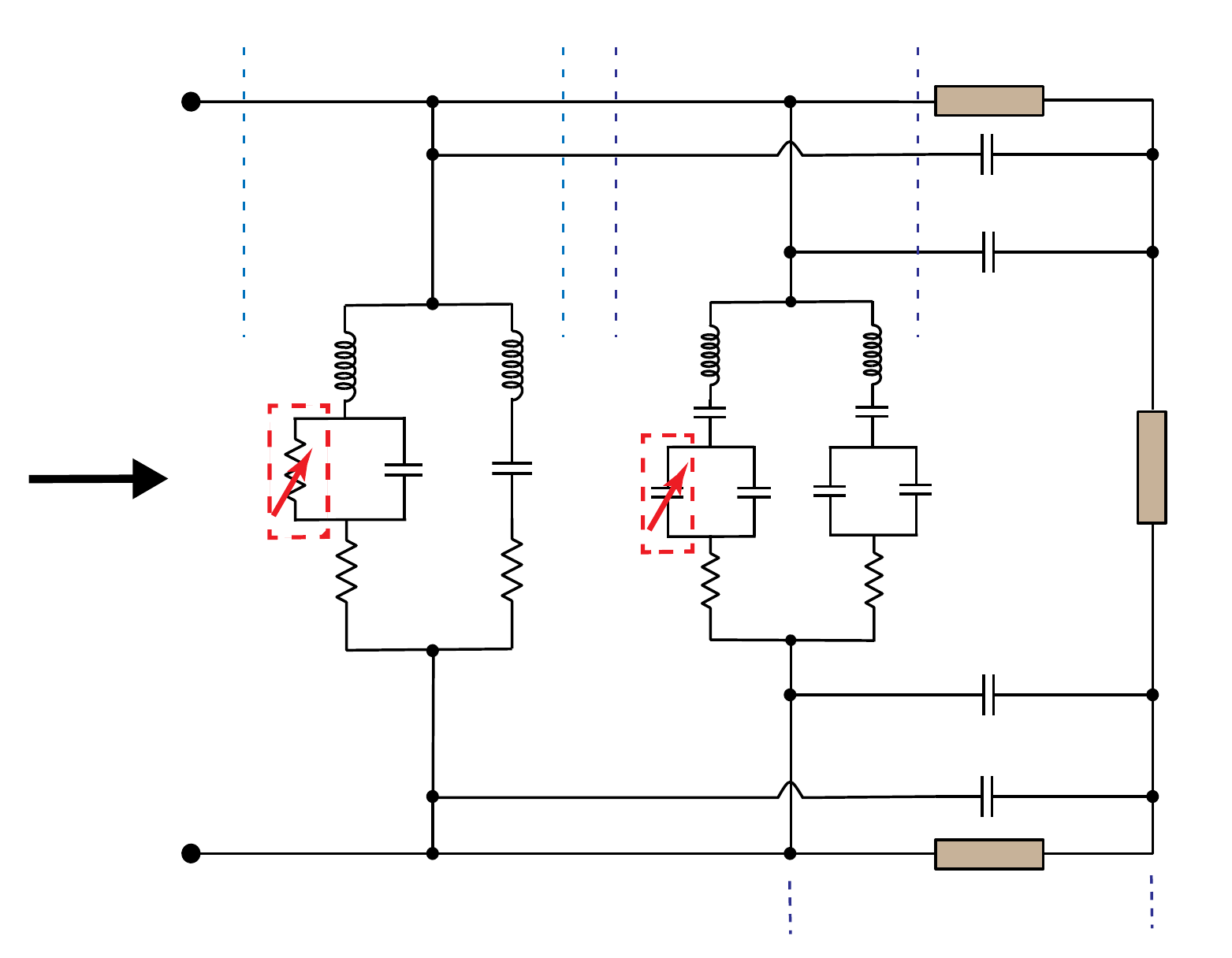}
			\put(24,47){\scriptsize $l_1$}
			\put(43,47){\scriptsize $l_2$}
			\put(35,39){\scriptsize $c_1$}
			\put(44,39){\scriptsize $c_2$}
			\put(23,30){\scriptsize $r_1$}
			\put(43,30){\scriptsize $r_2$}
			\put(78,61){\scriptsize $c_d$}
			\put(78,16){\scriptsize $c_d$}
			\put(18,39){\color{red}\fontsize{6}{4}$R$}
			
			\put(53,47){\scriptsize $l_3$}
			\put(72,47){\scriptsize $l_4$}
			\put(59,43){\scriptsize $c_3$}
			\put(65,43){\scriptsize $c_4$}
			\put(63,37){\scriptsize $c_5$}
			\put(69,37){\scriptsize $c_6$}
			\put(76,37){\fontsize{6}{4}$k$\color{red}$C$}
			\put(53,30){\scriptsize $r_3$}
			\put(72,30){\scriptsize $r_4$}
			\put(78,53){\scriptsize $c_s$}
			\put(78,24){\scriptsize $c_s$}
			\put(48.5,36.5){\color{red}\fontsize{6}{4}$C$}
			
			\put(-2,42){\fontsize{6}{4}$\Gamma(f_0),Z_0$}
			\put(77,72){\fontsize{6}{4}$Z, \theta$}
			\put(79,4){\makebox(0,0){\color{amber}\scriptsize \textbf{Grounded} Dielectric}}
			\put(96,46){\rotatebox{-90}{\fontsize{6}{4}$Z_\text{short}$}}
			\put(33,76){\makebox(0,0){\color{amber}\scriptsize \shortstack{DRR with Tunable \\ \textbf{Resistor}}}}
			\put(63,76){\makebox(0,0){\color{amber}\scriptsize \shortstack{SRR with Tunable \\ \textbf{Capacitor}}}}
		\end{overpic}
		\label{MS_ckt_model}}
	\subfigure[][]{
		\begin{overpic}[grid=false, width=0.9\columnwidth]{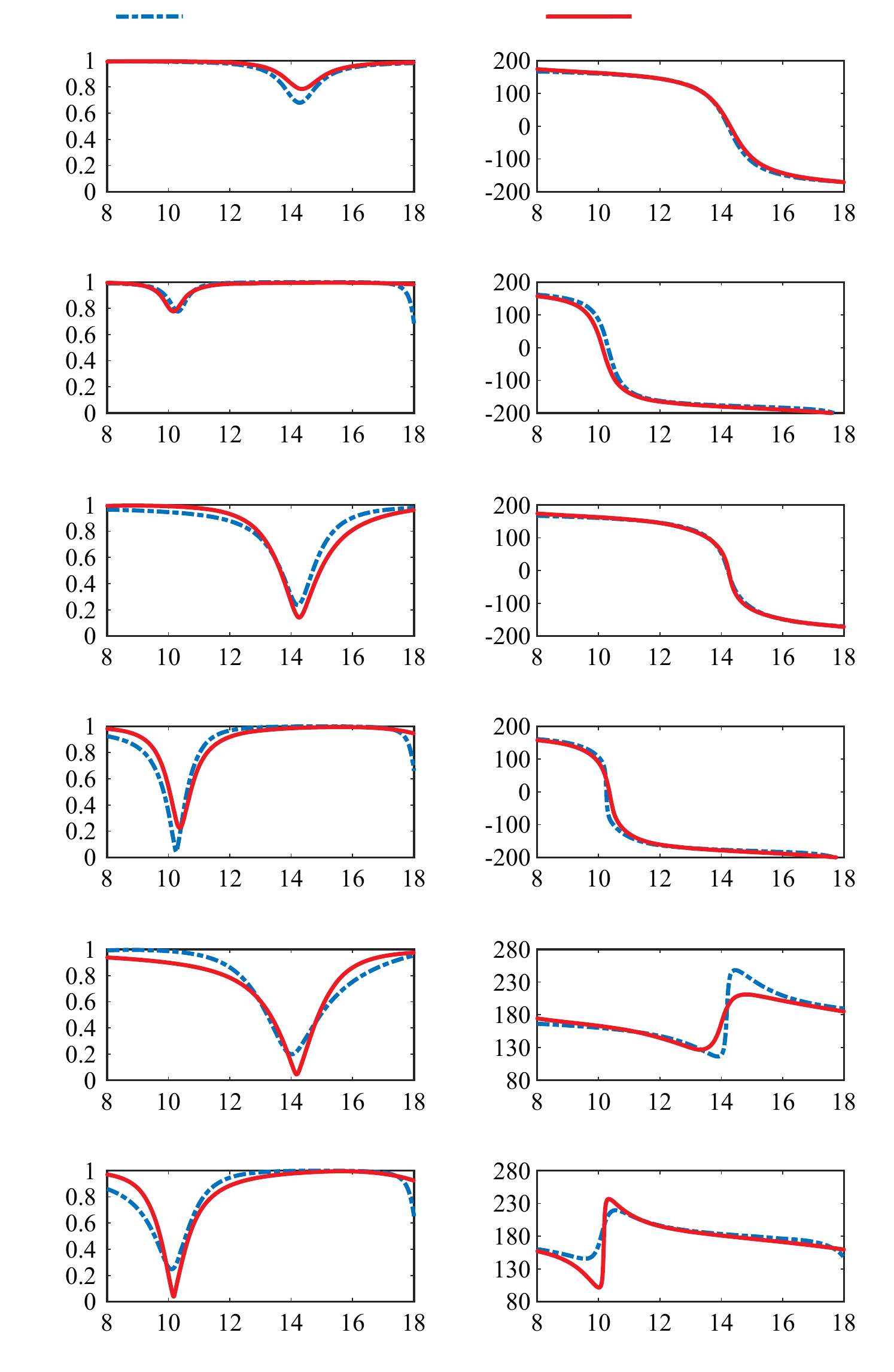}
			\put(13,1){\scriptsize Frequency (GHz)}
			\put(46,1){\scriptsize Frequency (GHz)}
			\put(2,4){\rotatebox{90}{\scriptsize Magnitude, $|\Gamma|$}}
			\put(34,4){\rotatebox{90}{\scriptsize Phase, $\angle \Gamma$ (deg)}}
			\put(65,10){\rotatebox{-90}{\makebox(0,0){\scriptsize \color{amber} \shortstack{$R=100~\Omega$ \\ $ C= 0.25~\text{pF}$}}}}

			\put(2,20){\rotatebox{90}{\scriptsize Magnitude, $|\Gamma|$}}
			
			\put(34,20){\rotatebox{90}{\scriptsize Phase, $\angle \Gamma$ (deg)}}
			\put(65,26){\rotatebox{-90}{\makebox(0,0){\scriptsize \color{amber} \shortstack{$R=100~\Omega$ \\ $ C= 0.025~\text{pF}$}}}}

			\put(2,36){\rotatebox{90}{\scriptsize Magnitude, $|\Gamma|$}}
			
			\put(34,36){\rotatebox{90}{\scriptsize Phase, $\angle \Gamma$ (deg)}}
			\put(65,42){\rotatebox{-90}{\makebox(0,0){\scriptsize \color{amber} \shortstack{$R=50~\Omega$ \\ $ C= 0.25~\text{pF}$}}}}
			
			\put(2,52){\rotatebox{90}{\scriptsize Magnitude, $|\Gamma|$}}
			
			\put(34,52){\rotatebox{90}{\scriptsize Phase, $\angle \Gamma$ (deg)}}
			\put(65,58){\rotatebox{-90}{\makebox(0,0){\scriptsize \color{amber} \shortstack{$R=50~\Omega$ \\ $ C= 0.025~\text{pF}$}}}}
			
			\put(2,69){\rotatebox{90}{\scriptsize Magnitude, $|\Gamma|$}}
			
			\put(34,69){\rotatebox{90}{\scriptsize Phase, $\angle \Gamma$ (deg)}}
			\put(65,75){\rotatebox{-90}{\makebox(0,0){\scriptsize \color{amber} \shortstack{$R=5~\Omega$ \\ $ C= 0.25~\text{pF}$}}}}
			
			\put(2,85){\rotatebox{90}{\scriptsize Magnitude, $|\Gamma|$}}
			
			\put(34,85){\rotatebox{90}{\scriptsize Phase, $\angle \Gamma$ (deg)}}
			\put(65,91){\rotatebox{-90}{\makebox(0,0){\scriptsize \color{amber} \shortstack{$R=5~\Omega$ \\ $ C= 0.025~\text{pF}$}}}}

			\put(14,97){\scriptsize \shortstack{Equivalent \\Circuit Model}}
			\put(48,97){\scriptsize \shortstack{Unit Cell\\FEM-HFSS}}
		\end{overpic}
		\label{hfss_vs_ckt_MS}}
	\caption{Proposed equivalent circuit model of the coupled-resonator based metasurface unit cell. (a) Equivalent circuit model. (b) Comparison of FEM-HFSS simulation with circuit model results for several $R$ and $C$ values. The various circuit's parameters are summarized in Tab.~\ref{full_wave_ckt_spec}.}
	\label{MS_ckt_model_hfss_vs_ckt}
\end{figure}
\begin{figure}[!t]
	\centering
	\subfigure[][]{
		\begin{overpic}[grid=false, width=0.9\columnwidth]{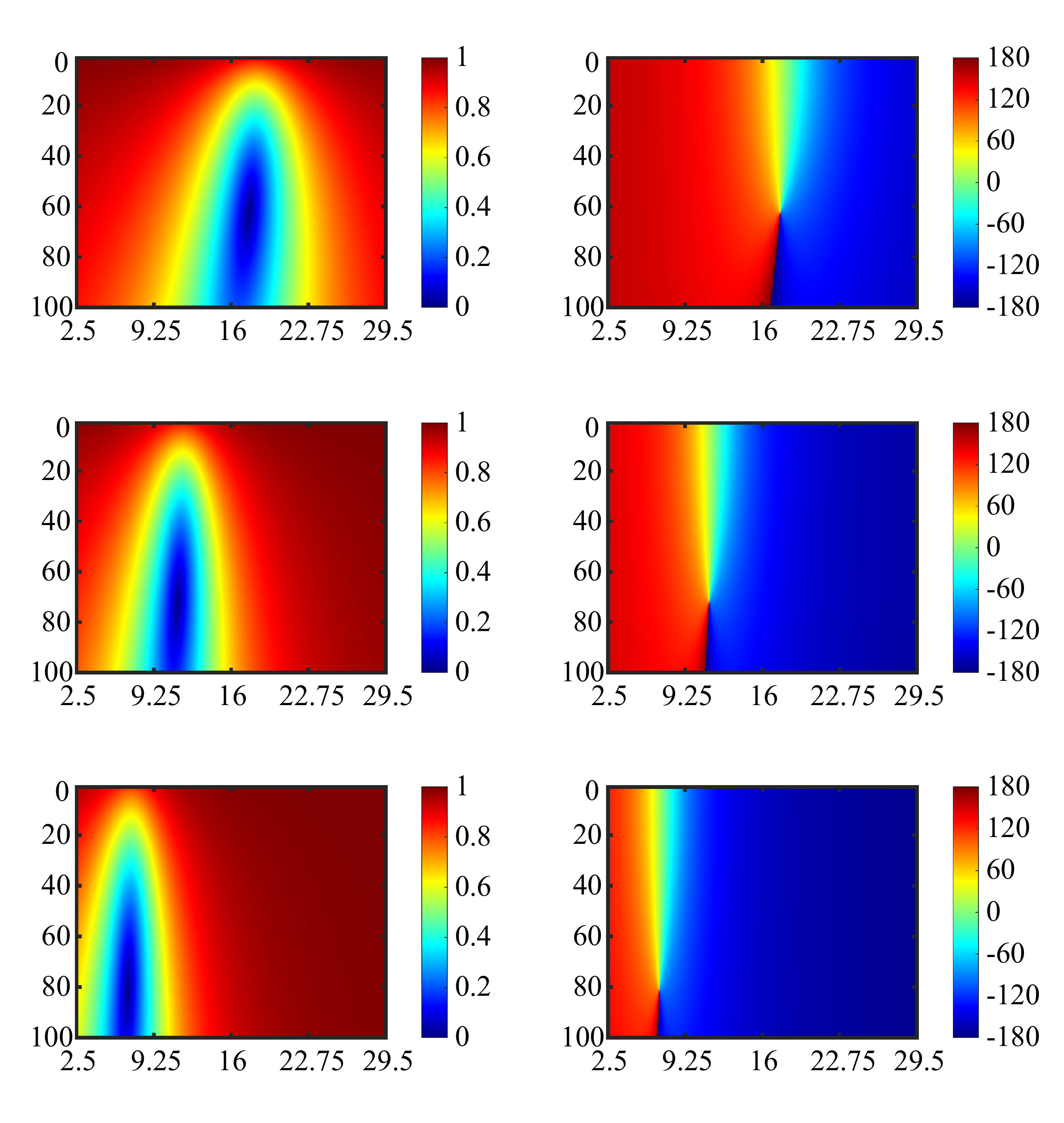}
			\put(7,1){\scriptsize Capacitance $\times10^{-2}$(pF)}
			\put(0,11){\rotatebox{90}{\scriptsize Resistance~($\Omega$)}}
			\put(55,1){\scriptsize Capacitance $\times10^{-2}$(pF)}
			\put(47,11){\rotatebox{90}{\scriptsize Resistance~($\Omega$)}}			
			\put(7,34){\scriptsize Capacitance $\times10^{-2}$(pF)}
			\put(0,44){\rotatebox{90}{\scriptsize Resistance~($\Omega$)}}
			\put(55,34){\scriptsize Capacitance $\times10^{-2}$(pF)}
			\put(47,44){\rotatebox{90}{\scriptsize Resistance~($\Omega$)}}			
			\put(7,67){\scriptsize Capacitance $\times10^{-2}$(pF)}
			\put(0,76){\rotatebox{90}{\scriptsize Resistance~($\Omega$)}}
			\put(55,67){\scriptsize Capacitance $\times10^{-2}$(pF)}
			\put(47,76){\rotatebox{90}{\scriptsize Resistance~($\Omega$)}}
			\put(20,98){\makebox(0,0){\scriptsize \color{matlabblue}\textsc{\textbf{Reflection Magnitude, $|\Gamma|$}}}}
			\put(70,98){\makebox(0,0){\scriptsize \color{matlabblue}\textsc{\textbf{Reflection Phase, $\angle \Gamma$}}}}
			\put(-4,20){\rotatebox{90}{\makebox(0,0){\scriptsize \color{amber} \boxed{\textsc{\textbf{$f_0 =13$~GHz}}}}}}
			\put(-4,52){\rotatebox{90}{\makebox(0,0){\scriptsize \color{amber} \boxed{\textsc{\textbf{$f_0 =12$~GHz}}}}}}
			\put(-4,84){\rotatebox{90}{\makebox(0,0){\scriptsize \color{amber} \boxed{\textsc{\textbf{$f_0 =11$~GHz}}}}}}
			\put(71,13){\makebox(0,0){\scriptsize \color{white} \textsc{\shortstack{Equivalent \\Circuit Model}}}}
		\end{overpic}
		\label{Amp_Phs_RC_two_resonators_ckt}}
	\subfigure[][]{
		\begin{overpic}[grid=false, width=0.9\columnwidth]{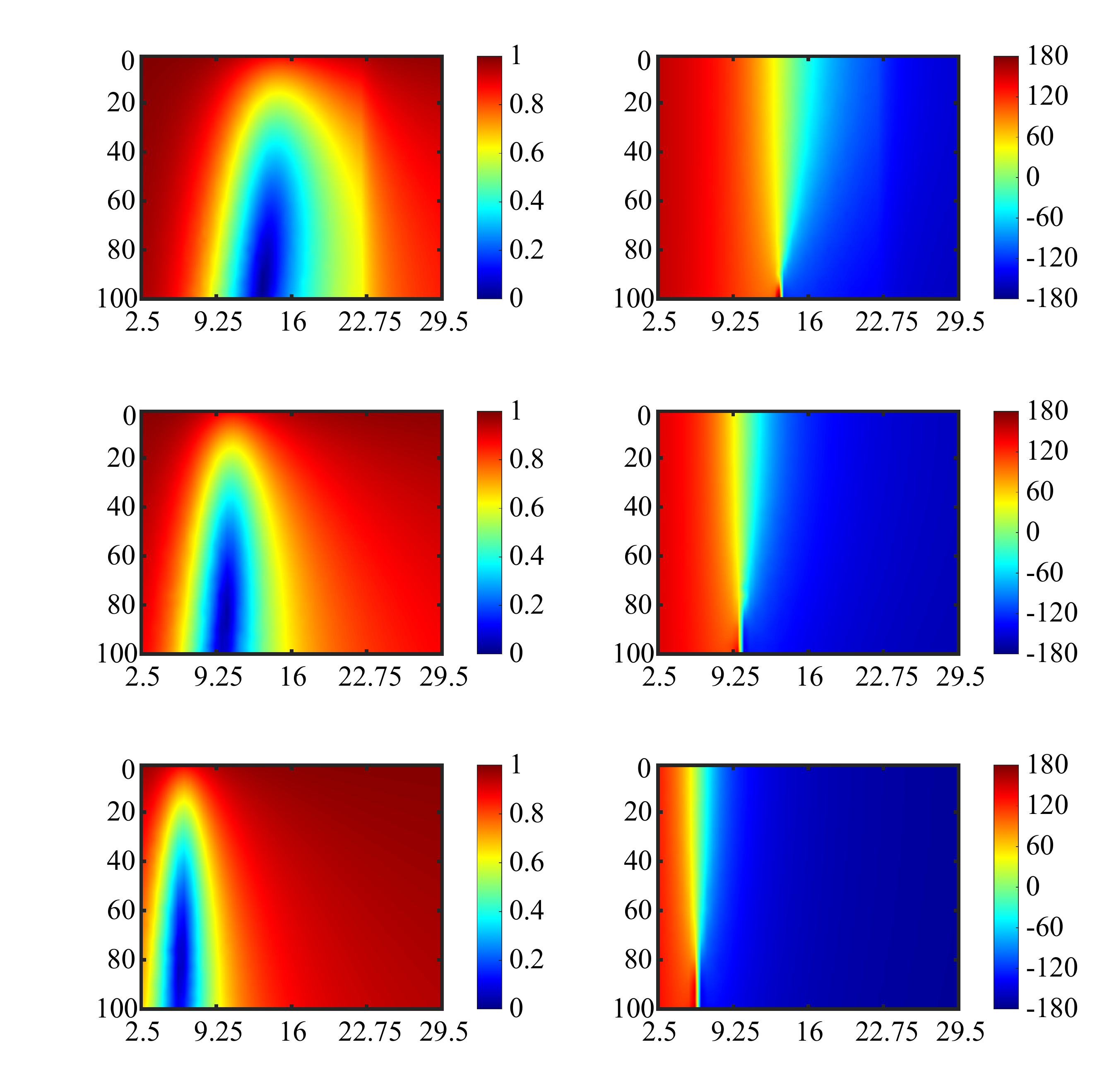}
			\put(7,1){\scriptsize Capacitance $\times10^{-2}$(pF)}
			\put(0,11){\rotatebox{90}{\scriptsize Resistance~($\Omega$)}}
			\put(55,1){\scriptsize Capacitance $\times10^{-2}$(pF)}
			\put(47,11){\rotatebox{90}{\scriptsize Resistance~($\Omega$)}}			
			\put(7,34){\scriptsize Capacitance $\times10^{-2}$(pF)}
			\put(0,44){\rotatebox{90}{\scriptsize Resistance~($\Omega$)}}
			\put(55,34){\scriptsize Capacitance $\times10^{-2}$(pF)}
			\put(47,44){\rotatebox{90}{\scriptsize Resistance~($\Omega$)}}			
			\put(7,67){\scriptsize Capacitance $\times10^{-2}$(pF)}
			\put(0,76){\rotatebox{90}{\scriptsize Resistance~($\Omega$)}}
			\put(55,67){\scriptsize Capacitance $\times10^{-2}$(pF)}
			\put(47,76){\rotatebox{90}{\scriptsize Resistance~($\Omega$)}}
			\put(20,98){\makebox(0,0){\scriptsize \color{matlabblue}\textsc{\textbf{Reflection Magnitude, $|\Gamma|$}}}}
			\put(70,98){\makebox(0,0){\scriptsize \color{matlabblue}\textsc{\textbf{Reflection Phase, $\angle \Gamma$}}}}
			\put(-4,20){\rotatebox{90}{\makebox(0,0){\scriptsize \color{amber} \boxed{\textsc{\textbf{$f_0 =13$~GHz}}}}}}
			\put(-4,52){\rotatebox{90}{\makebox(0,0){\scriptsize \color{amber} \boxed{\textsc{\textbf{$f_0 =12$~GHz}}}}}}
			\put(-4,84){\rotatebox{90}{\makebox(0,0){\scriptsize \color{amber} \boxed{\textsc{\textbf{$f_0 =11$~GHz}}}}}}
			\put(71,13){\makebox(0,0){\scriptsize \color{white} \textsc{\shortstack{FEM-HFSS}}}}
		\end{overpic}
		\label{Amp_Phs_RC_two_resonators_hfss}}
	\caption{Comparison of the equivalent circuit model (a) of Fig.~\ref{MS_ckt_model} with FEM-HFSS (b), shown using the $R-C$ dependent reflection amplitude-phase for the proposed coupled-resonator metasurface cell for capacitance values ranged between $0.025 - 0.295$~pF and resistance values ranged between $1 - 100$~$\Omega$ at different operating frequencies}
	\label{Amp_Phs_RC_two_resonators_ckt_hfss}
\end{figure}%
\section{Circuit Models Analysis}

A better insight into the proposed metasurface structure may be gained using an equivalent circuit model representation. Since the proposed metasurface's unit cell consists of two coupled-resonators, namely DRR and SRR, a circuit model is developed for each resonator individually with its respective tuning element. Then the overall response of the metasurface's unit cell is constructed by combining the developed equivalent circuit models of the two resonators, DRR and SRR, considering the electromagnetic coupling effects.

\subsection{Single Resonator (Dipole/Split-Ring)}
The two metasurface's unit cell resonators, i.e., the DRR and SRR resonators, can be modeled using a circuit model representation approach ~\cite{costa2014overview}. The equivalent circuit models are based on equivalent transmission lines representation of the metasurface's unit cell resonators. The DRR is modeled with a two-shunt combination of series $RLC$ when a normally incident free space plane-wave with a characteristic impedance of $Z_0$ excites the DRR with an E-field being polarized along the $y-$axis as shown in Fig.~\ref{dipole_ring_ckt_model}. When the DRR's exciting gap is loaded with an active element such as a PIN diode to control the overall impedance of the DRR, a loaded resistance $R$ representing the overall impedance of the active element is modeled in shunt across the DRR's gap $c_1$. The coupling between the adjacent cells is represented by $c_2$. 

The SRR equivalent circuit, on the other hand, is modeled similarly with a two-shunt combination of series $RLC$ as shown in Fig.~\ref{ring_ckt_model} with some modifications that account for the off-balanced normally incident E-field along the $y-$axis. An off-balanced E-field is seen by the two SRR's arms and between the adjacent cells when its gap $c_3$ is loaded with an active element such as a varactor $C$ while $c_1$ and $c_2$ represent the coupling between the adjacent cells on each SRR arms. Shunt capacitances of $c_4$ and $k(C)$ are modeled to account for cell asymmetry and account for the changes on the inter-element coupling capacitance due to the changes in the active loaded capacitance $C$. Finally, the grounded dielectric substrate on both the DRR and SRR resonators is modeled as a shorted transmission line with a characteristic impedance of $Z$ with shunt capacitance $c_d$ and $c_s$, respectively, accounting for the small electromagnetic coupling between the resonators and their grounded dielectric substrates. 

The equivalent circuit models of the two single resonators were simulated using the advanced design simulator (ADS) and compared with the full-wave finite element simulator (HFSS) for a wide range of frequency bandwidth. The full-wave response of the unit cells was curve fitted by numerically finding various lumped element values of the equivalent circuit model. Fig.~\ref{hfss_vs_ckt_dipole_n_ring} compares the full-wave simulation of the reflection's magnitude and phase with the obtained responses by the lumped element circuit models from 5$-$30~GHz for the DRR when its gap or its equivalent in the circuit model is loaded with a resistance of $R = 50$~$\Omega$, and the SRR when its gap or its equivalent in the circuit model is loaded with a capacitance of $C = 0.18$~pF. The full-wave simulation results of the two resonators show good agreement with their equivalent circuit models with the loaded elements $R$ and $C$ for large ranges of value (not shown here). This supports the suitability of the proposed circuit model to correctly represent the two physical models of DRR and SRR.

\subsection{coupled-resonator (Dipole-Split-Ring)}
The proposed metasurface unit cell based on coupled-resonators is next modeled using the above circuit models for the isolated resonators. The DRR and SRR circuit models above were superimposed to model the equivalent circuit model of the proposed metasurface unit cell. Fig.~\ref{MS_ckt_model} shows the superimposed configuration for the metasurface equivalent circuit model (its details lumped elements parameters are summarized in Tab.~\ref{full_wave_ckt_spec}. Once the two-unit cells are superimposed, they are electromagnetically coupled, which as a result perturbs its various circuit element values. Fig.~\ref{hfss_vs_ckt_MS} compares the reflection's magnitude and phase for the metasurface unit cell with the obtained responses by its equivalent circuit model using different values of $R$ and $C$ for the controlling loaded resistance and capacitance. The metasurface's circuit model magnitude and phase responses are in good agreement with the obtained full-wave simulation results. 

One major motivation of building an equivalent circuit model for the proposed metasurface (i.e; the coupled-resonator) is to build a faster approach than the full-wave simulation to examine the coverage range of magnitude and phase capability of the metasurface. Fig.~\ref{Amp_Phs_RC_two_resonators_ckt} shows two-dimensional contour views of the reflection's magnitude and phase coverage for the proposed metasurface unit cell obtained using its equivalent circuit model at three different frequencies. The used capacitance values $C$ are ranged between $0.025 - 0.295$~pF and the resistance values $R$ are ranged between $1 - 100$~$\Omega$ (typical among commercially available off-the-shelf varactor and PIN diodes). Similarly, the full-wave simulator HFSS is used to compare those contour coverages obtained by the equivalent circuit model at different operating frequencies as shown in Fig.~\ref{Amp_Phs_RC_two_resonators_hfss}. The contour views of the reflection magnitude and phase coverages show similar trends and frequency variations for results obtained using the equivalent circuit model and full-wave simulator. The results shown in Fig.~\ref{Amp_Phs_RC_two_resonators_ckt_hfss} thus not only substantiate the equivalent circuit model but also shows the capability of the proposed metasurface cells with the two combined resonators to have similar reflection magnitude and phase coverage at \emph{different operating frequencies}. 
\begin{table}
\renewcommand{\arraystretch}{1.2}
\caption{Full-Wave and Circuit Models Parameters}
\label{full_wave_ckt_spec}
\centering
\begin{tabular}{c c c}

    \textbf{ \shortstack{Resonator \\Structure}} & \textbf{\shortstack{Unit Cell \\Model~(mm)}} & \textbf{\shortstack{Circuit \\Model}} \\\hline

  \shortstack{Split Ring \\Resonator (SRR)}  &  \begin{tabular}[c]{@{}l@{}}$a = 2.92$\\ $b = 0.20$\\$l = 0.254$\end{tabular} & \begin{tabular}[c]{@{}l@{}}$l_{1} = 2.05$ $l_{2} = 12$~nH\\$c_{1} = 0.095$, $c_{2} = 0.031$~pF\\$c_{3} = 0.0749$, $c_{4} = 0.002$~pF\\ $c_{s} = 0.002$~pF,~$k=3.84$\\$r_{1} = 2.31$, $r_{2}=10~$~$\Omega$\end{tabular}\\
  \hline

  \shortstack{Dipole Ring \\Resonator (DRR)}   &  \begin{tabular}[c]{@{}l@{}}$c = 2.36$ \\$d =0.12$\\$e =0.35$ \\$g = 0.127$\end{tabular} & \begin{tabular}[c]{@{}l@{}}$l_{1} = 1.4025$, $l_{2} = 1.2258$~nH\\$c_{1} = 2.2258$, $c_{2} = 0.0169$~pF\\$c_{d} = 0.3042$~pF\\$r_{1} = 0.8$, $r_{2} = 0.2$~$\Omega$\end{tabular}\\
  \hline
   
  \shortstack{Coupled \\Resonators}   &  \begin{tabular}[c]{@{}l@{}} Same dimensions\\ as SRR/DRR\\ $\Lambda = 3$\\$s = 0.04$\\ $f = 0.08$\end{tabular} & \begin{tabular}[c]{@{}l@{}}$l_{1} = 2.2947$, $l_{2} = 2.2262$~nH\\$l_{3} = 0.8622$, $l_{4} = 0.7731$~nH\\ $c_{1} = 2.1972$, $c_{2} = 0.0473$~pF\\ $c_{3} = 0.2749$, $c_{4} = 0.001 $~pF\\$c_{5} = 0.08 $, $c_{6} = 5.2 $~pF\\$k=2.31$\\$c_{d} = 0.005$, $c_{s} = 0.0001$~pF\\ $r_{1} = 0.001 $, $r_{2} = 0.005$~$\Omega$\\$r_{3} = 0.0004$, $r_{4} = 2$~$\Omega$ \end{tabular}\\
   
    \hline
      
  &  \begin{tabular}[c]{@{}l@{}}Dielectric substrate\\(all resonators):\\RO3006 ($\epsilon_r = 6.15$\\$\tan\delta = 0.0024$\\thickness of $0.64$)\end{tabular} & \begin{tabular}[c]{@{}l@{}} All circuit models:\\ Free-space impedance:\\ $Z_0 = 377$~$\Omega$\\ TL impedance:\\$Z = 152$~$\Omega$, \\$\theta = 34.28^\circ$ at 18~GHz \end{tabular}\\ \hline
\end{tabular}
\end{table}%
\section{Full-Wave Demonstration}
\begin{figure}[htbp]
	\centering
	\begin{overpic}[grid=false, width=0.9\columnwidth]{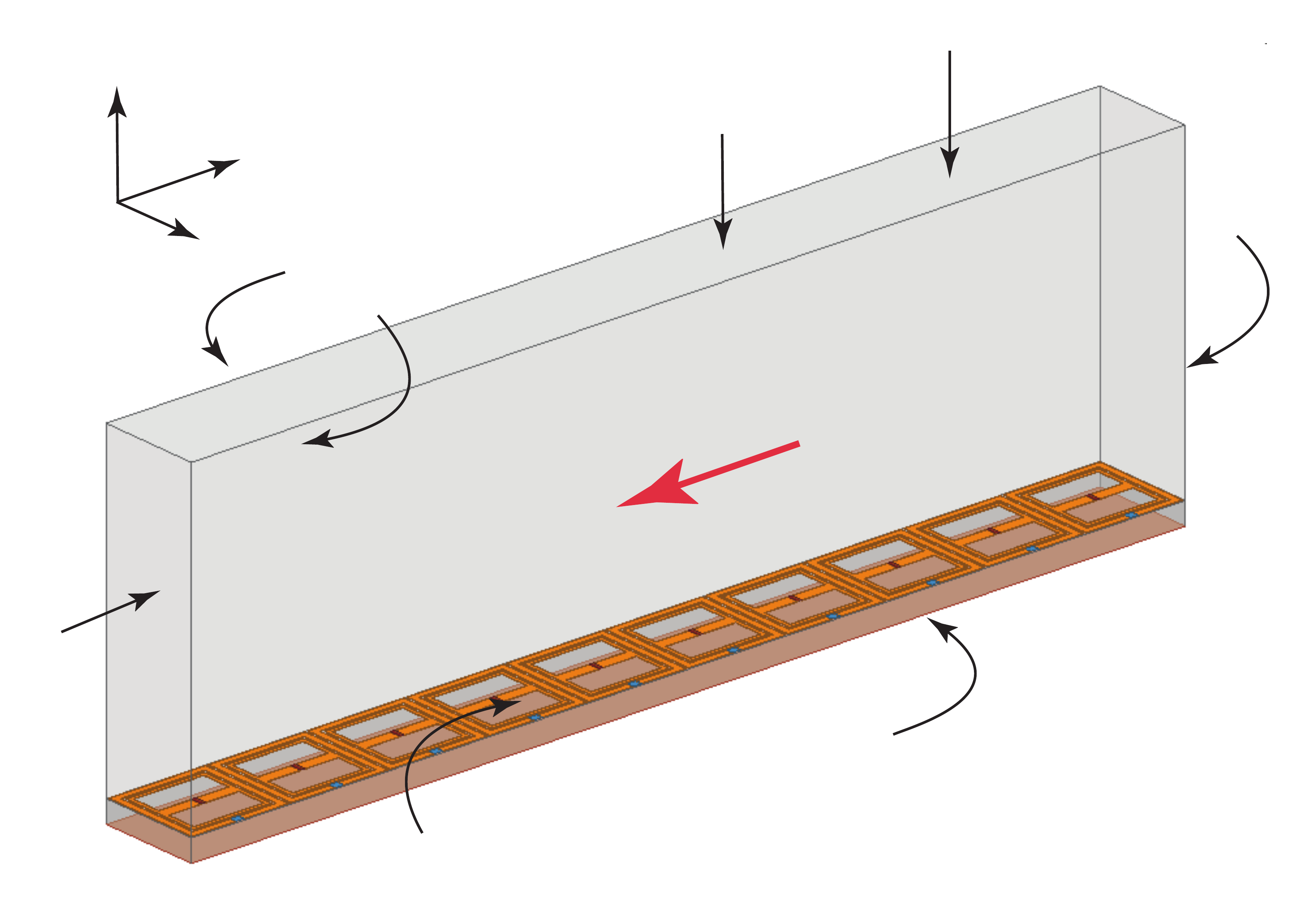}
		\put(-7,19){\scriptsize \shortstack{Radiation \\Boundary}}
		\put(92,53){\scriptsize \shortstack{Radiation \\Boundary}}
		\put(67,66){\scriptsize \shortstack{Radiation \\Boundary}}
		\put(23,47){\scriptsize PMC}
		\put(57,11){\scriptsize \shortstack{PEC \\ reflector}}
		\put(25,3){\scriptsize Metasurface, $N$ cells}
		\put(48,60){\scriptsize \shortstack{Uniform \\Plane-wave}}
		\put(44,31){\makebox(0,0){\scriptsize $E$}}
		\put(20,58){\makebox(0,0){\scriptsize $y$}}
		\put(9,65){\makebox(0,0){\scriptsize $z$}}
		\put(17,50){\makebox(0,0){\scriptsize $x$}}
	\end{overpic}
	\caption{Illustration of the simulation setup in FEM-HFSS for a finite size metasurface consisting of $N$ unit cells with $N$ lumped resistors and capacitors. The unit cell array is excited with a normally incident linearly polarized plane wave.}
	\label{ms_sim_set_presp}
\end{figure}

This section will demonstrate different numerical examples examining the reflection capability of the proposed metasurface to meet specific beam-forming, tilting, and splitting specifications, using simultaneous amplitude and phase control obtained through the integrated active resistors and capacitors. The HFSS simulation setup to model a coupled array of identical unit cells forming the metasurface is illustrated in Fig~\ref{ms_sim_set_presp} (for reasonable computational time). The amplitude and variation are obtained by varying the lumped elements (and not the geometry) across the surface. The finite size metasurface is assumed to be excited by a normally incident plane-wave along the $z-$axis with an $E-$field polarized along the $y-$axis, and where the metasurface structure is assumed to be uniform along the $x-$axis using PMC and PEC boundaries, respectively. Radiation boundaries enclose the structure top and sides. A fixed operation frequency of 12.5~GHz is assumed in all illustration examples.
\begin{figure*}[htbp]
	\centering
	\subfigure[][]{
	\begin{overpic}[grid=false, width=0.95\columnwidth]{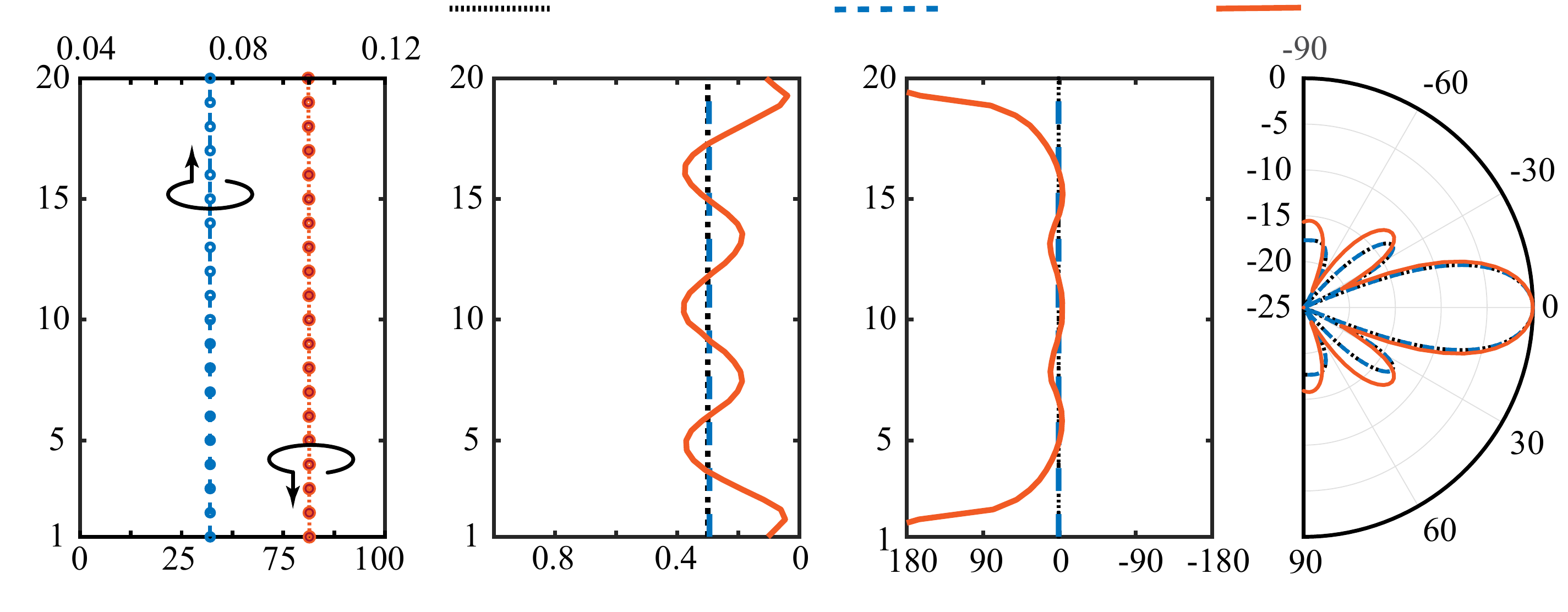}
	\put(8,0){\tiny Resistance~($\Omega$)}
	\put(8,38){\tiny Capacitance~(pF)}
   	\put(0,13){\rotatebox{90}{\tiny Cell number, $n$}}
   	\put(37,0){\tiny Magnitude}  	
   	\put(64,0){\tiny Phase~(deg)}  	
   	\put(36,38){\tiny AF (theoretical)}
   	\put(61,38){\tiny Unit cell (HFSS)}
    	\put(84,38){\tiny Metasurface (HFSS)}
   	\put(15,45){\makebox(0,0){\scriptsize \color{matlabblue} \shortstack{\textsc{\textbf{Lumped Element}}\\$C(n),~R(n)$}}}
	\put(42,45){\makebox(0,0){\scriptsize \color{matlabblue}\shortstack{\textsc{\textbf{Near-field }} \\ $|E(y,z_0)|$}}}
	\put(67,45){\makebox(0,0){\scriptsize \color{matlabblue}\shortstack{\textsc{\textbf{Near-Field }} \\$\angle E(y,z_0)$}}}
	\put(90,45){\makebox(0,0){\scriptsize \color{matlabblue}\shortstack{\textsc{\textbf{Far-field Gain}} \\$G_\theta(\theta,\phi=90^\circ)$}}}
	\put(-4,20){\rotatebox{90}{\makebox(0,0){\scriptsize \color{amber} \boxed{\textsc{\textbf{$\theta_\text{peak}=0^\circ$, $|\Gamma_1|$}}}}}}
	\end{overpic}
	\label{uniform_cell_at_mag_03_08_A}}\hfill
	\subfigure[][]{
	\begin{overpic}[grid=false, width=0.95\columnwidth]{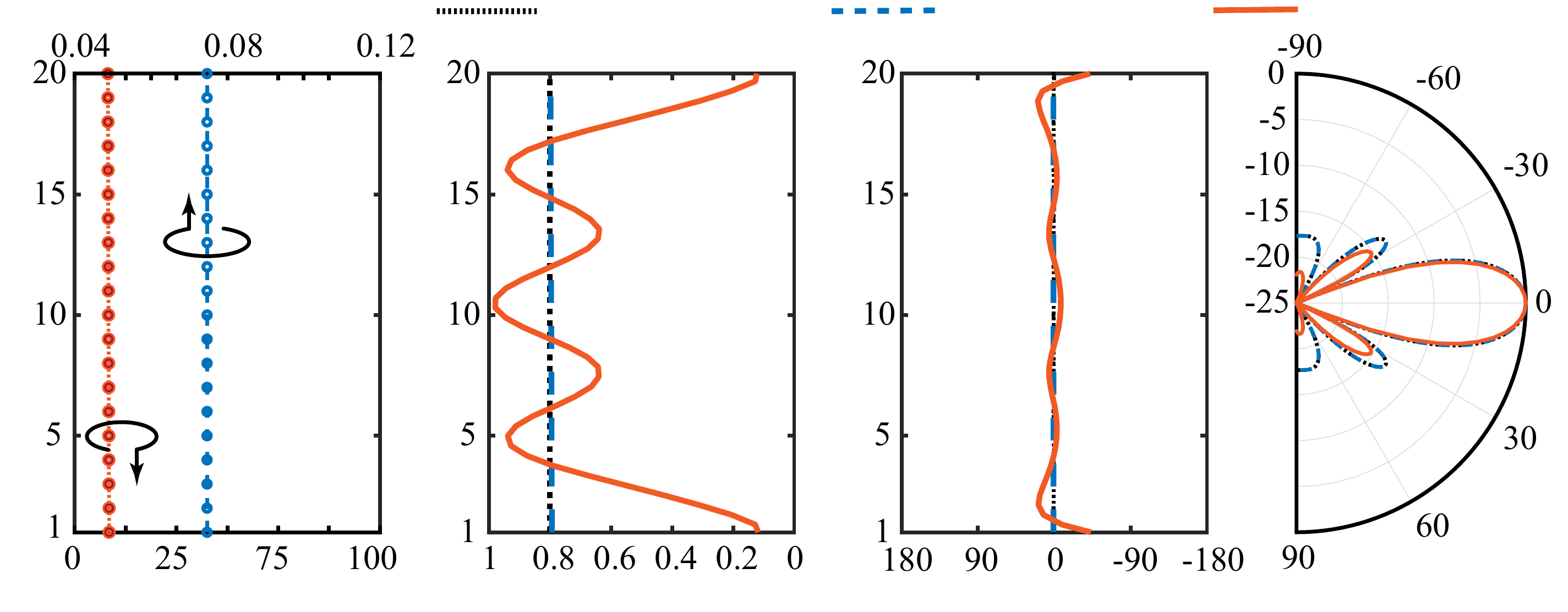}
	\put(8,0){\tiny Resistance~($\Omega$)}
	\put(8,38){\tiny Capacitance~(pF)}
   	\put(0,13){\rotatebox{90}{\tiny Cell number, $n$}}
   	\put(37,0){\tiny Magnitude}  	
   	\put(64,0){\tiny Phase~(deg)}  	
   	\put(36,38){\tiny AF (theoretical)}
   	\put(61,38){\tiny Unit cell (HFSS)}
    	\put(84,38){\tiny Metasurface (HFSS)}
   	\put(15,45){\makebox(0,0){\scriptsize \color{matlabblue} \shortstack{\textsc{\textbf{Lumped Element}}\\$C(n),~R(n)$}}}
	\put(42,45){\makebox(0,0){\scriptsize \color{matlabblue}\shortstack{\textsc{\textbf{Near-field }} \\ $|E(y,z_0)|$}}}
	\put(67,45){\makebox(0,0){\scriptsize \color{matlabblue}\shortstack{\textsc{\textbf{Near-Field }} \\$\angle E(y,z_0)$}}}
	\put(90,45){\makebox(0,0){\scriptsize \color{matlabblue}\shortstack{\textsc{\textbf{Far-field Gain}} \\$G_\theta(\theta,\phi=90^\circ)$}}}
	\put(104,20){\rotatebox{-90}{\makebox(0,0){\scriptsize \color{amber} \boxed{\textsc{\textbf{$\theta_\text{peak}=0^\circ$, $|\Gamma_2|$}}}}}}
	\end{overpic}
	\label{uniform_cell_at_mag_03_08_B}}\hfill
	\subfigure[][]{
	\begin{overpic}[grid=false, width=0.95\columnwidth]{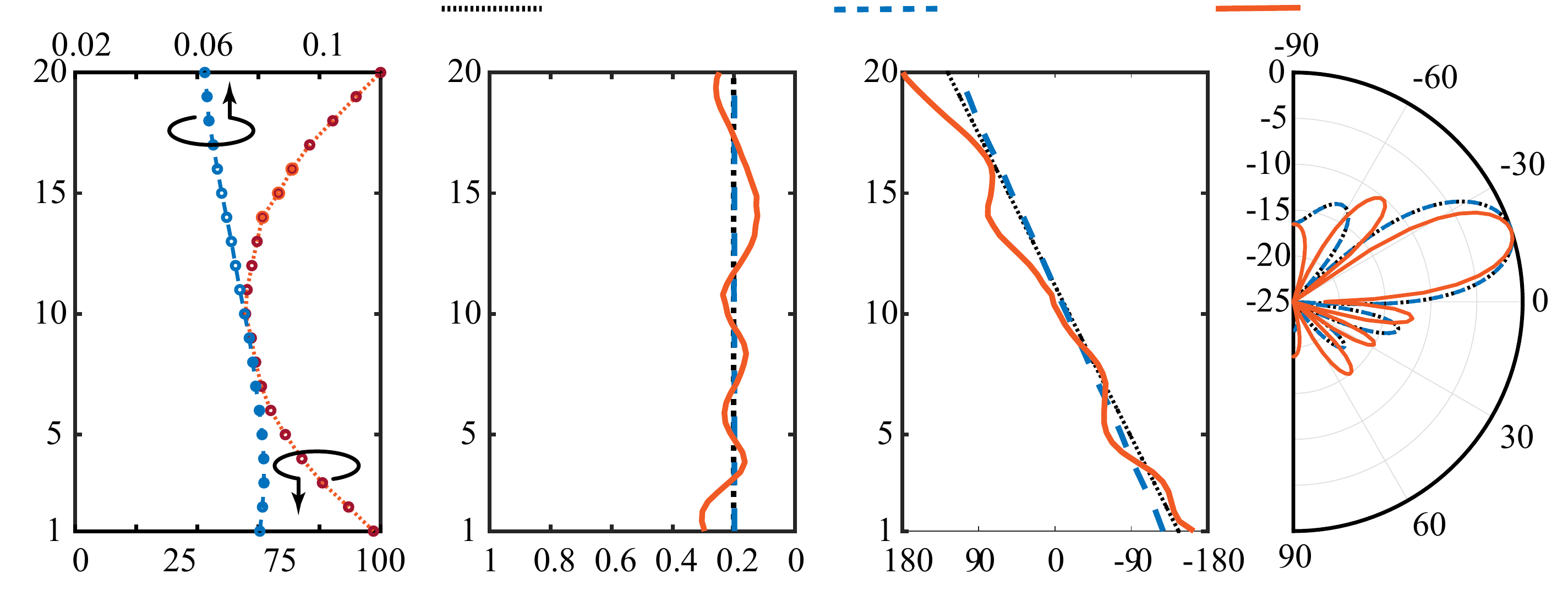}
	\put(8,0){\tiny Resistance~($\Omega$)}
	\put(8,38){\tiny Capacitance~(pF)}
   	\put(0,13){\rotatebox{90}{\tiny Cell number, $n$}}
   	\put(37,0){\tiny Magnitude}  	
   	\put(64,0){\tiny Phase~(deg)}  	
   	\put(36,38){\tiny AF (theoretical)}
   	\put(61,38){\tiny Unit cell (HFSS)}
    	\put(84,38){\tiny Metasurface (HFSS)}
	\put(-4,20){\rotatebox{90}{\makebox(0,0){\scriptsize \color{amber} \boxed{\textsc{\textbf{$\theta_\text{peak}=-15^\circ$, $|\Gamma_1|$}}}}}}
	\end{overpic}
	\label{const_mag_02_lin_phs_A}}\hfill
	\subfigure[][]{
	\begin{overpic}[grid=false, width=0.95\columnwidth]{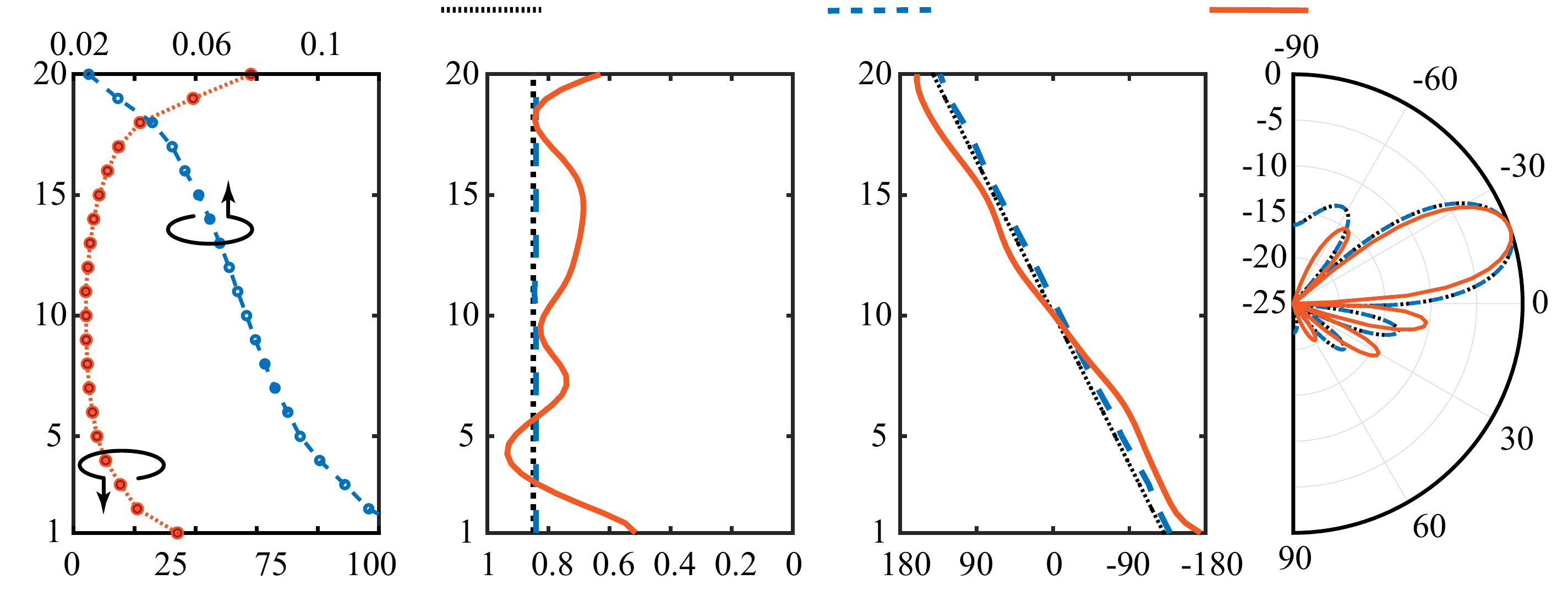}
	\put(8,0){\tiny Resistance~($\Omega$)}
	\put(8,38){\tiny Capacitance~(pF)}
   	\put(0,13){\rotatebox{90}{\tiny Cell number, $n$}}
   	\put(37,0){\tiny Magnitude}  	
   	\put(64,0){\tiny Phase~(deg)}  	
   	\put(36,38){\tiny AF (theoretical)}
   	\put(61,38){\tiny Unit cell (HFSS)}
    	\put(84,38){\tiny Metasurface (HFSS)}
	\put(104,20){\rotatebox{-90}{\makebox(0,0){\scriptsize \color{amber} \boxed{\textsc{\textbf{ $\theta_\text{peak}=-15^\circ$, $|\Gamma_2|$}}}}}}
	\end{overpic}
	\label{const_mag_085_lin_phs_B}}\hfill
	\caption{Metasurface for beam-deflection with variable gain control. (a)-(b) Specified peak reflection of $0.3$ and $0.8$, respectively, with no beam deflection. (c)-(d) Specified peak reflection of $0.2$ and $0.85$, respectively, with beam deflection to $15^\circ$. Normally incident plane-wave is assumed in all cases and metasurface consists of 20 cells. Each plot shows the resistance, $R$ and capacitance $C$ variation across the surfaces, along with spatially varying magnitude and phase. Array Factor (AF) shows the ideal element distributions, ``unit cell" corresponds to a physical unit cell simulated in HFSS with Floquet boundary conditions, and ``metasurface" corresponds to a non-uniform full-wave structure with varying $R$ and $C$. The reflected scattered fields are shown in the far-field using radiation pattern plots.}	
	\label{uniform_cell_at_mag_03_08_const_mag_085_lin_phs_02_085}
\end{figure*}
It's well-known that having linear array elements of uniform magnitudes with constant phase progression will produce a maximum reflection beam directed broadside to the axis of the excited linear array elements. Thus, we start examining the reflection response of metasurface structure imposing uniform magnitudes and constant phase progression on the metasurface's cells to meet such specifications. Two examples with different uniform field magnitudes of 0.3 and 0.8 were considered with a constant phase. Towards this approach, an array factor pattern of a $20-$element linear array with an inter-element spacing of $\lambda/8$ is assumed as a required specification to produce two broadside reflection gains (i.e.; uniform elements magnitude of 0.3 and 0.8 with constant phase progression). The choice of such amplitudes is strictly used to illustrate the gain reflection capability for the metasurface while other amplitude values can be assumed. The inter-element spacing is $\lambda/8$ that is strictly determined based on the proposed unit cell size at the operating frequency of 12.5~GHz. Thus, the array elements specifications to meet the broadside reflection requirement are: $N =$~20, $d =$~$\lambda/8$, $\beta =$~0, and $a_n =$~0.3 and 0.8.

Next, appropriate values of the resistors and capacitors were chosen (using the lookup tables similar to the contour plots of Fig.~\ref{Amp_Phs_RC_two_resonators_ckt_hfss}) in the physical unit cell of the coupled-resonator configuration, to achieve the desired reflection. The complex reflectance of each unit cell is next used in the array factor to compute the analytical far-fields (equivalent to uncoupled arrays, and labeled as ``Unit Cell (HFSS)"), as shown in Fig.~\ref{uniform_cell_at_mag_03_08_A} and \ref{uniform_cell_at_mag_03_08_B}. Finally, a finite-sized 1-D array of metasurfaces was built following the model of Fig.~\ref{ms_sim_set_presp}, and the computed full-wave response is superimposed with the AF specifications and the full-wave unit cell model (labeled as ``Metasurface (HFSS)"). The measured reflection phase and the magnitude shows ripples around the desired values due to stronger element couplings near the structure, and with dropping phase and amplitude near the two edges due to finite structure size. Despite these small deviations, a very good agreement between the far-field AF specifications, physical unit cell phase AF, and the finite-sized metasurface structure is obtained.
\begin{figure}[!t]
	\centering
	\subfigure[][]{
	\begin{overpic}[grid=false, width=0.45\columnwidth]{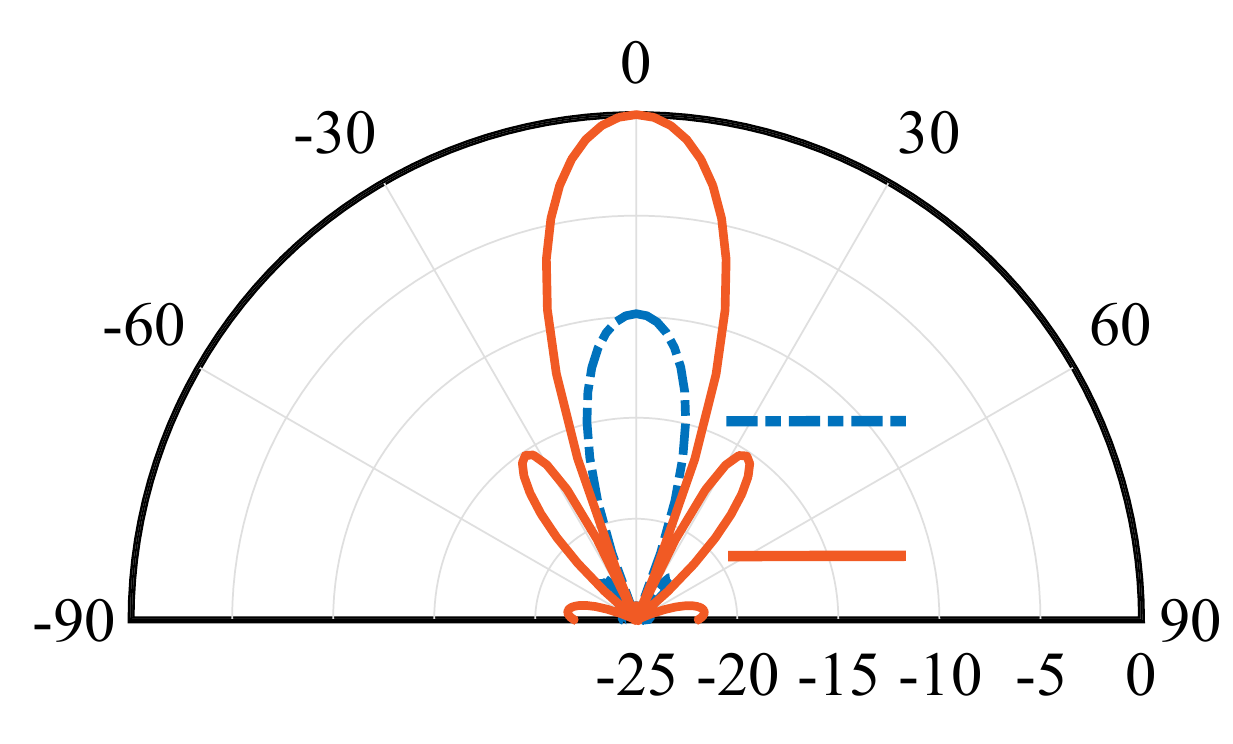}
		\put(75,25){\fontsize{3}{1}$\abs{S_{11}}=0.3$}
		\put(75,14){\fontsize{3}{1}$\abs{S_{11}}=0.8$}
		 \put(55,62){\makebox(0,0){\scriptsize \color{amber} \shortstack{\textsc{\textbf{No Tilt}}, Variable Gain}}}
	\end{overpic}
		\label{AF_polar_uniform_cell_at_mag_03_08}}\hfill
	\subfigure[][]{
	\begin{overpic}[grid=false, width=0.45\columnwidth]{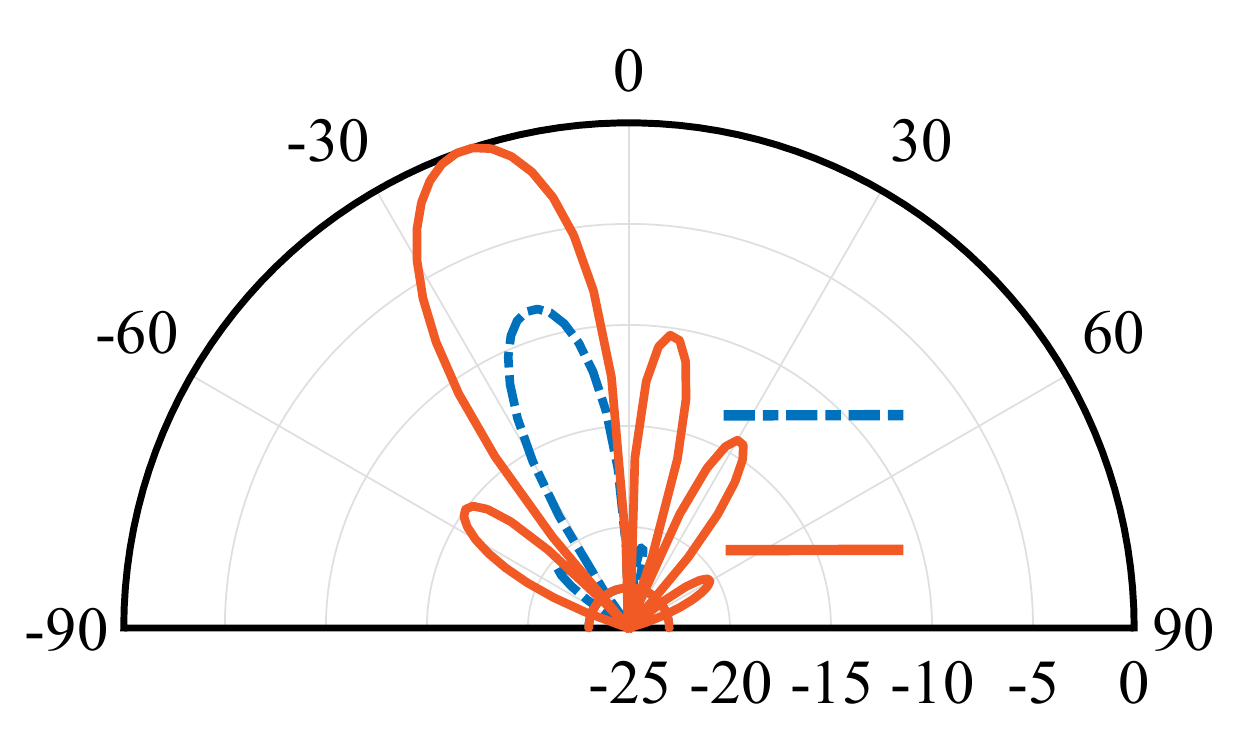}
		\put(75,25){\fontsize{3}{1}$\abs{S_{11}}=0.2$}
		\put(75,14){\fontsize{3}{1}$\abs{S_{11}}=0.85$}
		 \put(55,62){\makebox(0,0){\scriptsize \color{amber} \shortstack{\textsc{\textbf{Beam Tilt}}, Variable Gain}}}
		\end{overpic}
		\label{AF_polar_const_mag_02_085_lin_phs}}
	\caption{Normalized far-field radiation patterns obtained by the FEM-HFSS for a finite metasurface of $20$ coupled cells for: (a) a constant phase progression with two uniform field magnitudes of $0.3$ and $0.8$, and (b) a linear phase progressions with different field magnitudes of $0.2$ and $0.85$.}
	\label{polar_uniform_cell_at_mag_03_08_&_polar_const_mag_02_085_lin_phs}
\end{figure}
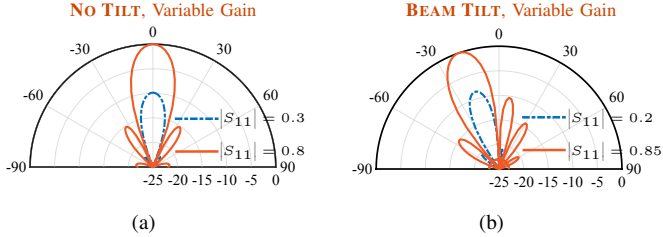
A further step was then taken from the above two mentioned examples that only showed a constant phase progression with uniform magnitudes. The metasurface reflector, in the following two examples, is studied to meet a requirement of having a linear phase progression where the beam is tilted with two different uniform magnitudes. The proposed required specifications, in this case, are to have the maximum of the array factor of the uniform linear array with a beam-tilting angle of $-15^{\circ}$ from the broadside to the axis of the array, while controlling the reflection (0.2 and 0.85, respectively). A similar approach to the above examples was then taken to extract the resistance and capacitance values, and to compute array factors for the specifications, coupled and uncoupled arrays. Figs.~\ref{const_mag_02_lin_phs_A} and~\ref{const_mag_085_lin_phs_B} shows the array factor patterns of a $20-$element uniform broadside array of two maximum reflection of magnitude of 0.2 and 0.85, respectively, where $N =$~20, $d =$~$\lambda/8$, $\beta =$~0.2, $a_n =$~0.2 and 0.85. This time, since a uniform magnitude and non-uniform phase distribution is desired, the resistance and capacitance values, $R_n$ and $C_n$ must vary across the surface. They are consequently chosen using the lookup tables similar to Fig.~\ref{Amp_Phs_RC_two_resonators_ckt_hfss} for the operating frequency of 12.5~GHz. The lumped elements variation is also shown in Fig.~\ref{uniform_cell_at_mag_03_08_const_mag_085_lin_phs_02_085} for all cases, which automatically makes the metasurface, electrically non-uniform. The corresponding near-field and far-field responses are shown in Fig.~\ref{uniform_cell_at_mag_03_08_const_mag_085_lin_phs_02_085} show a very good agreement with specifications, whereby the desired tilt in the beam with the desired gain is obtained.

Fig.~\ref{AF_polar_uniform_cell_at_mag_03_08} further Illustrates the controlling of the reflection gain of the metasurface for the above two examples comparing the far-field radiation patterns for the two uniform field magnitudes of 0.3 and 0.8 with a constant phase and Fig.~\ref{AF_polar_const_mag_02_085_lin_phs} compares the refection gains for the latter two examples with the linear phase progression cases where the beam was tilted to an angle of $-15^{\circ}$ from broadside. Clearly, the resistance control in the DRR provides gain tuning in the far-field, which could range from a near-perfect reflection (except accounting for dissipation losses) to perfect absorption.

\begin{figure}[!t]
	\centering
	\subfigure[][]{
	\begin{overpic}[grid=false, width=0.95\columnwidth]{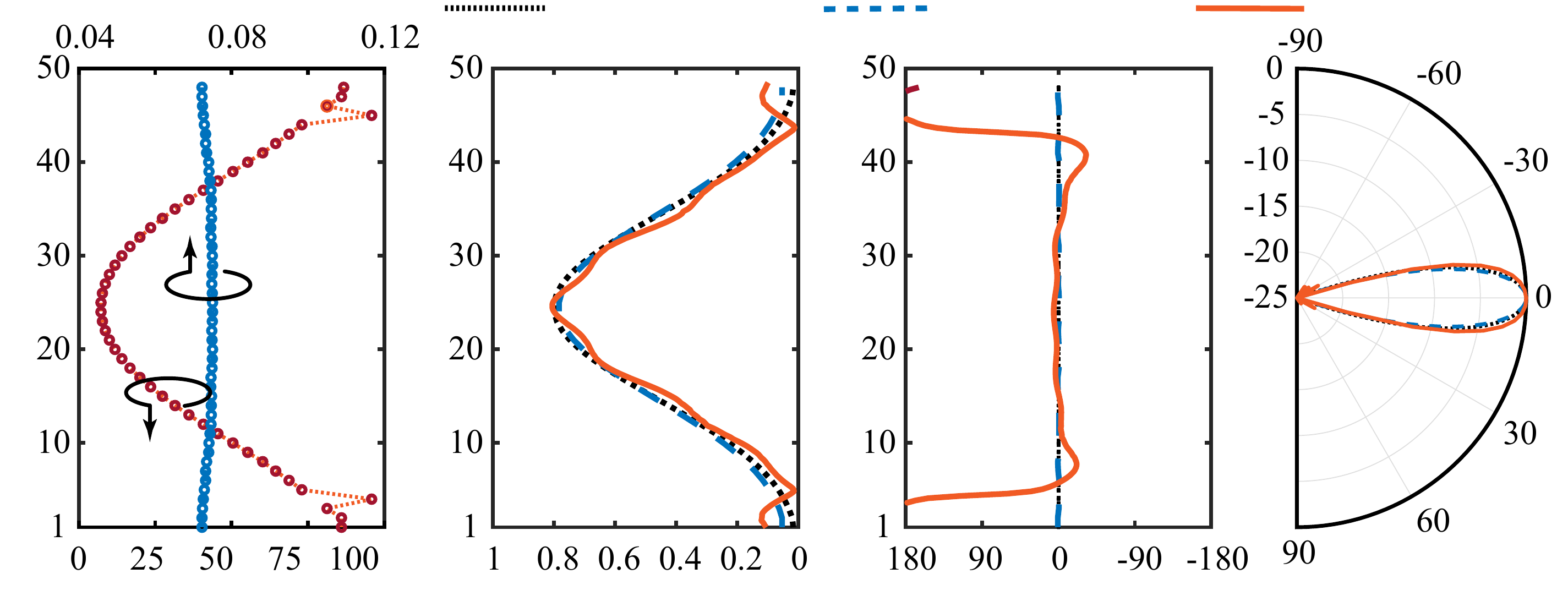}
	\put(8,0){\tiny Resistance~($\Omega$)}
	\put(8,38){\tiny Capacitance~(pF)}
   	\put(0,13){\rotatebox{90}{\tiny Cell number, $n$}}
   	\put(37,0){\tiny Magnitude}  	
   	\put(64,0){\tiny Phase~(deg)}  	
   	\put(36,38){\tiny AF (theoretical)}
   	\put(61,38){\tiny Unit cell (HFSS)}
    	\put(84,38){\tiny Metasurface (HFSS)}
   	\put(15,45){\makebox(0,0){\scriptsize \color{matlabblue} \shortstack{\textsc{\textbf{Lumped Element}}\\$C(n),~R(n)$}}}
	\put(42,45){\makebox(0,0){\scriptsize \color{matlabblue}\shortstack{\textsc{\textbf{Near-field }} \\ $|E(y,z_0)|$}}}
	\put(67,45){\makebox(0,0){\scriptsize \color{matlabblue}\shortstack{\textsc{\textbf{Near-Field }} \\$\angle E(y,z_0)$}}}
	\put(90,45){\makebox(0,0){\scriptsize \color{matlabblue}\shortstack{\textsc{\textbf{Far-field Gain}} \\$G_\theta(\theta,\phi=90^\circ)$}}}
	\put(-4,20){\rotatebox{90}{\makebox(0,0){\scriptsize \color{amber} \boxed{\textsc{\textbf{Broadside Reflection}}}}}}
	\end{overpic}
		\label{cheby_mag_const_phase_01}}\hfill
	\subfigure[][]{
	\begin{overpic}[grid=false, width=0.95\columnwidth]{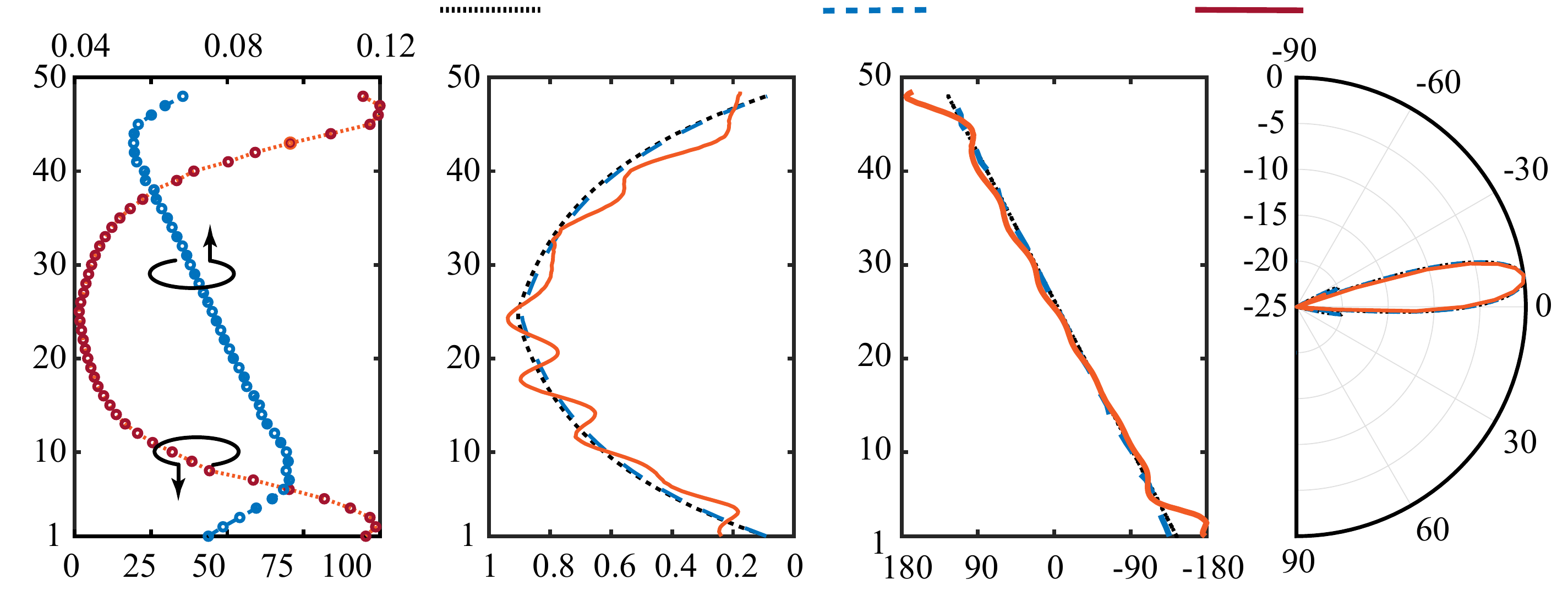}
	\put(8,0){\tiny Resistance~($\Omega$)}
	\put(8,38){\tiny Capacitance~(pF)}
   	\put(0,13){\rotatebox{90}{\tiny Cell number, $n$}}
   	\put(37,0){\tiny Magnitude}  	
   	\put(64,0){\tiny Phase~(deg)}  	
   	\put(36,38){\tiny AF (theoretical)}
   	\put(61,38){\tiny Unit cell (HFSS)}
    	\put(84,38){\tiny Metasurface (HFSS)}
	\put(-4,20){\rotatebox{90}{\makebox(0,0){\scriptsize \color{amber} \boxed{\textsc{\textbf{Off-Broadside Reflection}}}}}}
	\end{overpic}
		\label{Aprox_Binomial_magn_lin_phs_01}}
			\caption{Metasurface reflector for beam-deflection with controlled side-lobes (minimum 25~dB below peak gain). (a) Chebyshev magnitude distribution and a constant phase progression (no beam deflection) (b) Approximated binomial magnitude distribution and a linear phase progression (beam deflection). Normally incident plane-wave is assumed, and the metasurface is 48 cells long.}
	\label{cheby_mag_const_phase_01_Aprox_Binomial_magn_lin_phs_01}
\end{figure}
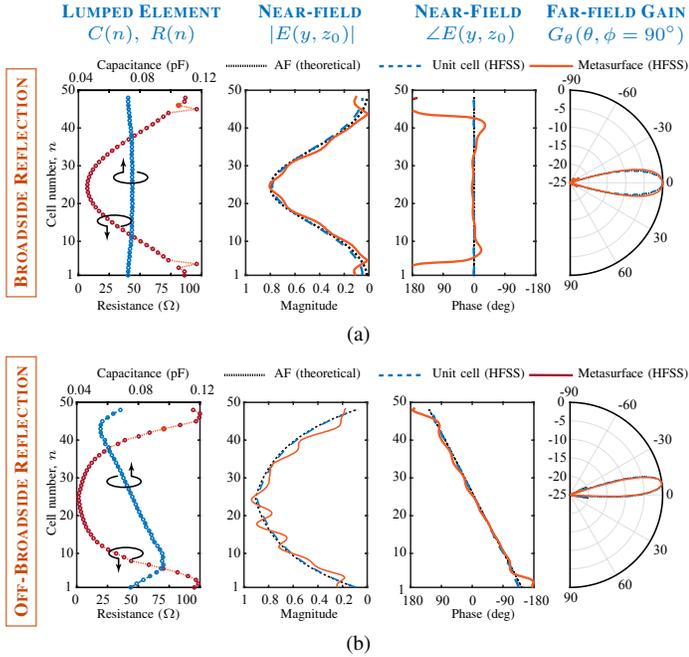

Next, we imposed more sophisticated scenarios where we assumed a specific side lobes level (grating lobes) that is required for broadside reflection as well as for achieving beam deflection. Thus, the following two examples will require the metasurface to maintain a lower sider level and to exhibit non-uniform magnitudes with constant and linear phase progression, respectively. The first example in the category is to have a reflection in a broadside direction. This example assumes that the specifications are having maximum side lobes at least $25$~dB below the main lobe directed along broadside while the main beam width is as small as possible. To meet these specifications, all elements will have the same phase excitation (i.e.; constant progression phase) and non-uniform amplitude excitation of the array elements. For the element amplitude, a Chebyshev array of $N =$~48 elements and inter-element spacing of $d =$~$\lambda/8$ is chosen to meet these specifications, as Chebyshev profiles are well-known to provide equi-ripple side-lobes in antenna theory. The Chebyshev AF of $N-$element array requires a Chebyshev polynomial $T_m(z)$ of $m=N-1$ order that is defined as follow:
	\begin{equation}
		T_m(z)=
	\begin{cases}
		(-1)^m \cosh\{m.\cosh^{-1} |z|\}, z\leqslant -1\\
		\cos\{m\cos^{-1}(z)\},~-1\leqslant z\leqslant 1\\
		\cosh\{m\cosh^{-1} (z)\}, z\geq 1
	\end{cases}
	\label{Eq:Cheby_poly}
  \end{equation}
\noindent and the ratio $R_0$ of major to minor lob intensity is the maximum of $T_{N-1}$ that is fixed at an argument $z_0$ ($|z_0|>1$) where $T_m^\text{max}(z_0)= R_0$. The specified AF using the Chebyshev polynomial is then equated to determining the coefficients for each power of $z$ that satisfies an $R_0$ of $25$~dB for a broadside beam with $\beta = 0$. Complex weights are obtained for the array element excitations. Figs.~\ref{cheby_mag_const_phase_01} shows an example of this Chebyshev linear array with a broadside beam and compares the required AF specifications with both the unit cell and the metasurface. A near-perfect magnitude and phase response are observed, and again, Despite a constant phase, both $R$ and $C$ are varied across the surface. Consequently, an excellent match between the realized pattern and the specifications is observed in the far-field.
	
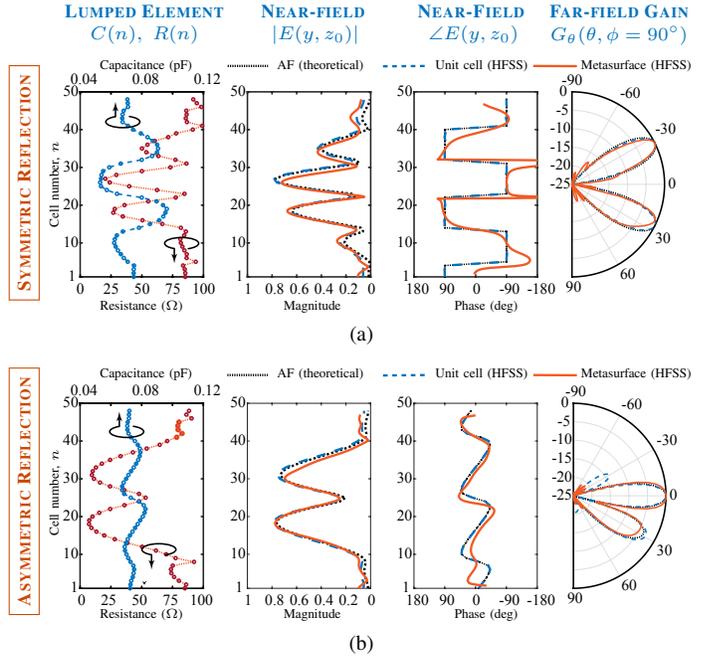
\begin{figure}[!t]
	\centering
	\subfigure[][]{
	\begin{overpic}[grid=false, width=0.95\columnwidth]{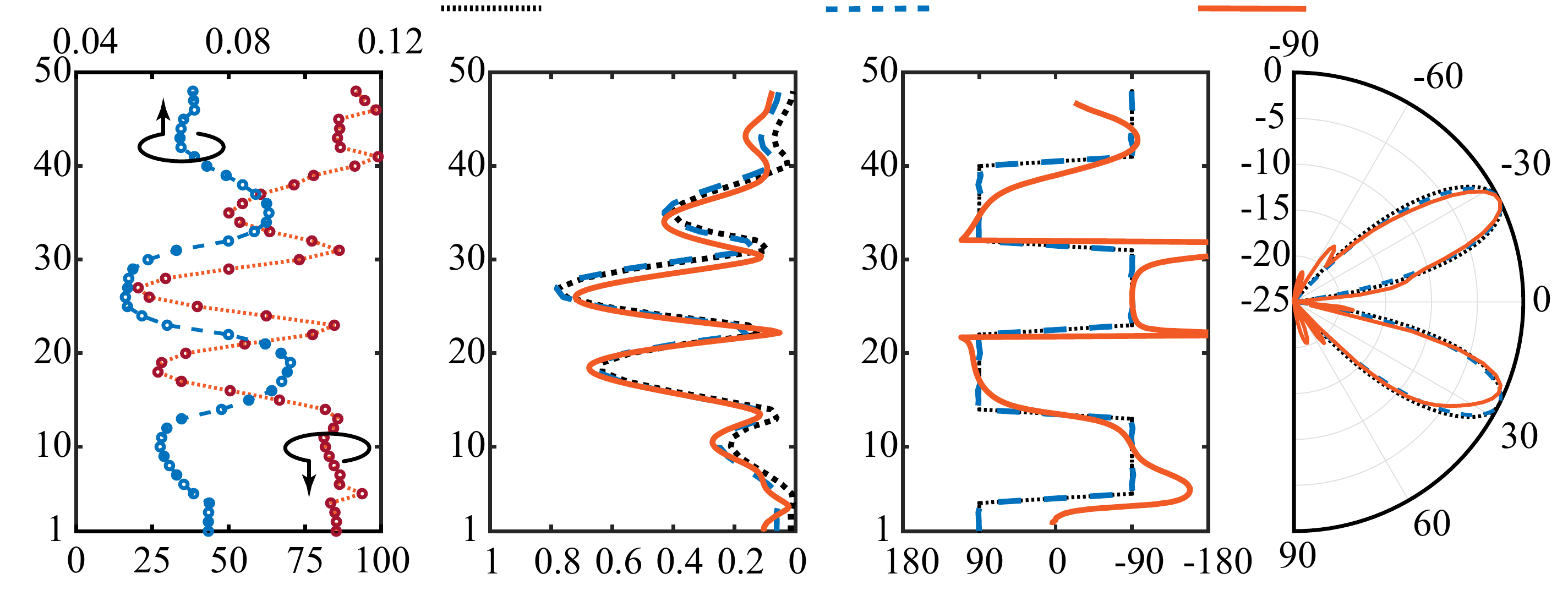}
	\put(8,0){\tiny Resistance~($\Omega$)}
	\put(8,38){\tiny Capacitance~(pF)}
   	\put(0,13){\rotatebox{90}{\tiny Cell number, $n$}}
   	\put(37,0){\tiny Magnitude}  	
   	\put(64,0){\tiny Phase~(deg)}  	
   	\put(36,38){\tiny AF (theoretical)}
   	\put(61,38){\tiny Unit cell (HFSS)}
    	\put(84,38){\tiny Metasurface (HFSS)}
   	\put(15,45){\makebox(0,0){\scriptsize \color{matlabblue} \shortstack{\textsc{\textbf{Lumped Element}}\\$C(n),~R(n)$}}}
	\put(42,45){\makebox(0,0){\scriptsize \color{matlabblue}\shortstack{\textsc{\textbf{Near-field }} \\ $|E(y,z_0)|$}}}
	\put(67,45){\makebox(0,0){\scriptsize \color{matlabblue}\shortstack{\textsc{\textbf{Near-Field }} \\$\angle E(y,z_0)$}}}
	\put(90,45){\makebox(0,0){\scriptsize \color{matlabblue}\shortstack{\textsc{\textbf{Far-field Gain}} \\$G_\theta(\theta,\phi=90^\circ)$}}}
	\put(-4,20){\rotatebox{90}{\makebox(0,0){\scriptsize \color{amber} \boxed{\textsc{\textbf{Symmetric Reflection}}}}}}
	\end{overpic}
		\label{Cheby_two_beams_01}}
		
			\subfigure[][]{
	\begin{overpic}[grid=false, width=0.95\columnwidth]{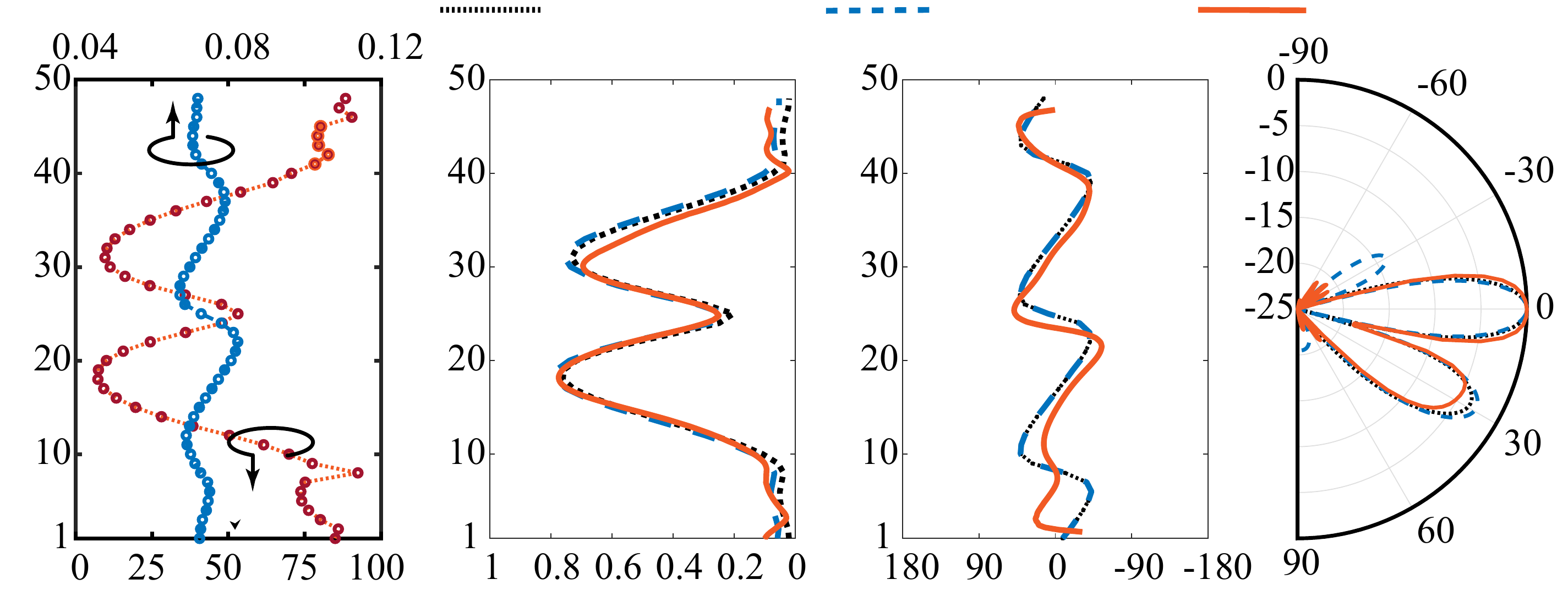}
	\put(8,0){\tiny Resistance~($\Omega$)}
	\put(8,38){\tiny Capacitance~(pF)}
   	\put(0,13){\rotatebox{90}{\tiny Cell number, $n$}}
   	\put(37,0){\tiny Magnitude}  	
   	\put(64,0){\tiny Phase~(deg)}  	
   	\put(36,38){\tiny AF (theoretical)}
   	\put(61,38){\tiny Unit cell (HFSS)}
    	\put(84,38){\tiny Metasurface (HFSS)}
	\put(-4,20){\rotatebox{90}{\makebox(0,0){\scriptsize \color{amber} \boxed{\textsc{\textbf{Asymmetric Reflection}}}}}}
	\end{overpic}
		\label{Cheby_two_beams_02}}
	\caption{Metasurface reflector for dual-beam scattering. (a) Two symmetric beams with respect to broadside with equal peak magnitudes, with reduced side-lobe levels following Chebyshev magnitude distribution. (b) Two asymmetric beams with reduced side-lobe levels following Chebyshev magnitude distribution and unequal peak magnitudes. Normally incident plane-wave is assumed, and the metasurface is 48 cells long.}
	\label{Cheby_two_beams}
\end{figure}

Then, similar to the above specifications for having maximum side lobes with at least a $25$~dB below the main lobe of an array factor, we also added another requirement - that is to tilt the reflected beam with an angle of 7$^{\circ}$ from the broadside, for instance. Thus, we investigated non-standard magnitude distributions, mainly to see how well the surface will follow such distributions, for the array elements' magnitude that is in a form of an approximate binomial distribution that is defined as follow:
	\begin{equation}
	a_n =
	\begin{cases}
		1-0.9^n, &  \begin{tabular}{l}
	              for $n=1~to~N/2$
				 \end{tabular}\\
		1-0.9^{\{n-(2m+1)\}}, & \begin{tabular}{ll}
						for $n=(N/2+1)~to~N$;\\
					       $m=0~to~(N/2-1)$
					       \end{tabular}
	\end{cases}
		\label{Eq:approx_binomial}
	\end{equation}
\noindent A linear array with $48-$element with an inter-element spacing of $d =$~$\lambda/8$ and non-uniform amplitudes is again used (i.e.; the approximate binomial expressed on Eq.~\ref{Eq:approx_binomial}). An array factor pattern satisfying those specifications is obtained for a $48-$element array ($N =$~48, $d =$~$\lambda/8$, $\beta =$~0.1~rad/m). Fig.~\ref{Aprox_Binomial_magn_lin_phs_01} shows and compares the required specifications' AF with both the unit cell (i.e.; capacitances and resistances obtained from the uncoupled metasurface unit cell) and the finite linear metasurface array. Again, a clear beam tilt with low side-lobes is obtained as a result of simultaneous variation of phase and magnitude across the surface.
Lastly, we examine the metasurface reflector by specifying two scenarios for having obtaining multi-beams in reflection. The first specification requirement is to have a two-beam reflection with identical gains where one beam is tilted at 27$^{\circ}$ from the broadside and the other one is at -27$^{\circ}$ with a major-to-minor lobe ratio of $R_0 = 60$~dB. The Chebyshev array is used here to meet such specifications. The two beams are equated with linear phase progression of ($\beta = 0.35$~rad/m) and ($\beta = -0.35$~rad/m) to satisfies the required tilting angles of $\pm27^{\circ}$, respectively. Figs.~\ref{Cheby_two_beams_01} shows the resulting two identically directive beams with minimum side lobes with an identical gain compared to the specifications, showing an excellent agreement. The second specification requirement is a more general one, with two beams having different gains where the higher gain beam is on the broadside and the other beam is at a 30$^{\circ}$ angle from the broadside with major-to-minor lobe ratio of $R_0 = 40$~dB, i.e. two beams with asymmetry. The broadside beam with the higher amplitude weights beam is having a constant progression phase with $\beta = 0$, and the tilted beam with 30$^{\circ}$ from the broadside is having a linear phase progression with $\beta = -0.38$~rad/m. The resulting two-beams with minimum side lobes with different reflection gains are shown in Figs.~\ref{Cheby_two_beams_02} comparing them to the required specifications, and showing very good agreement. This case thus clearly exemplifies the requirement of simultaneous magnitude and phase control as such an asymmetric beam cannot be easily realized using either amplitude or phase control only.

\section{Discussion}
It is noticed from the contour views shown in Fig.~\ref{Amp_Phs_RC_two_resonators_ckt_hfss}, that the proposed coupled-resonator structure of a fixed unit cell period $\Lambda$, is capable to operate at different frequencies. Each operating frequency requires a different range of resistance and capacitance to provide a maximized coverage of the reflection's amplitude $|\Gamma(f_0)|\in [0,~1]$ and phase $\angle \Gamma(f_0)\in[0,~2\pi]$ independently. This brings the importance of discussing various features of the proposed unit cell expressing the impact of the unit cell size on the realized magnitude and phase. 

The above full-wave illustration examples of having real-time independent magnitude and phase control for the metasurface reflector were chosen to operate at a frequency of 12.5~GHz. The proposed metasurface unit cell is $\lambda/8$ at this operating frequency that shows independent reflection magnitude and phase coverages as shown in Figs.~\ref{Amp_Phs_RC_two_resonators} and \ref{Amp_Phs_RC_two_resonators_unique} with capacitance and resistance ranged between $0.025 - 0.295$~pF and $1 - 100$~$\Omega$, respectively. While this range of capacitance, in particular, can cover a wide range of frequencies as suggested in Fig.~\ref{Amp_Phs_RC_two_resonators_ckt_hfss}, one may choose to have a smaller range of capacitance to operate at a single operating frequency while maintaining and not compromising the reflection and phase coverages (for greater tuning sensitivity, for instance). For example, capacitance values ranged between $0.025 - 0.16$~pF are sufficiently enough when operating at a frequency of 12.5~GHz to get the possible independent reflection magnitude and phase (refer to Fig.~\ref{Amp_Phs_RC_two_resonators}). Increasing the capacitance range will only produce replicates of similar independent magnitude and phase points (refer to monochromatic color on the contour view shown in Fig.~\ref{Amp_Phs_RC_two_resonators} as well as to the redundant concentrate number of points near the lower-right region of Fig.~\ref{Amp_Phs_RC_two_resonators_unique}). 
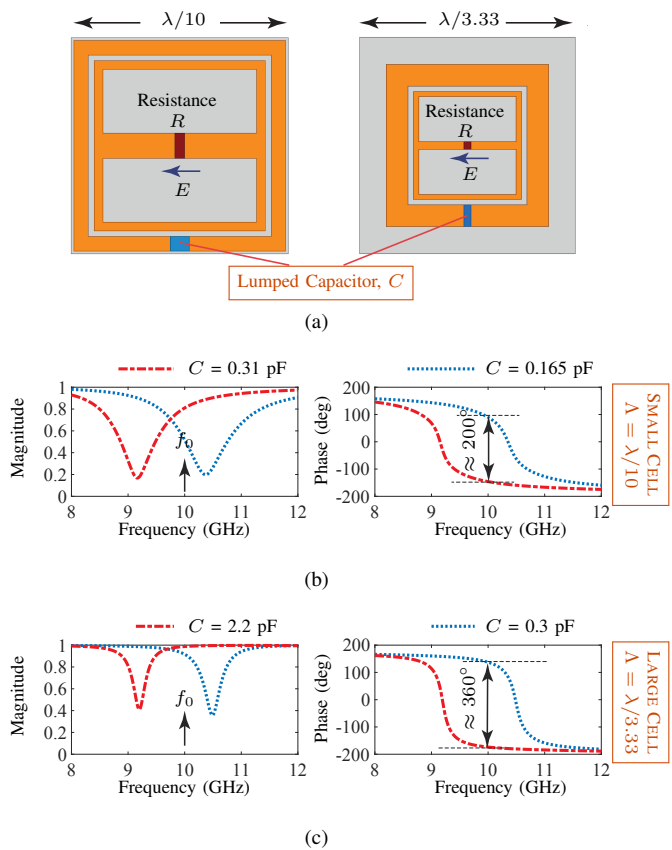
\begin{figure}[htbp]
	\centering
		\subfigure[][]{
	\begin{overpic}[grid=false, width=0.8\columnwidth]{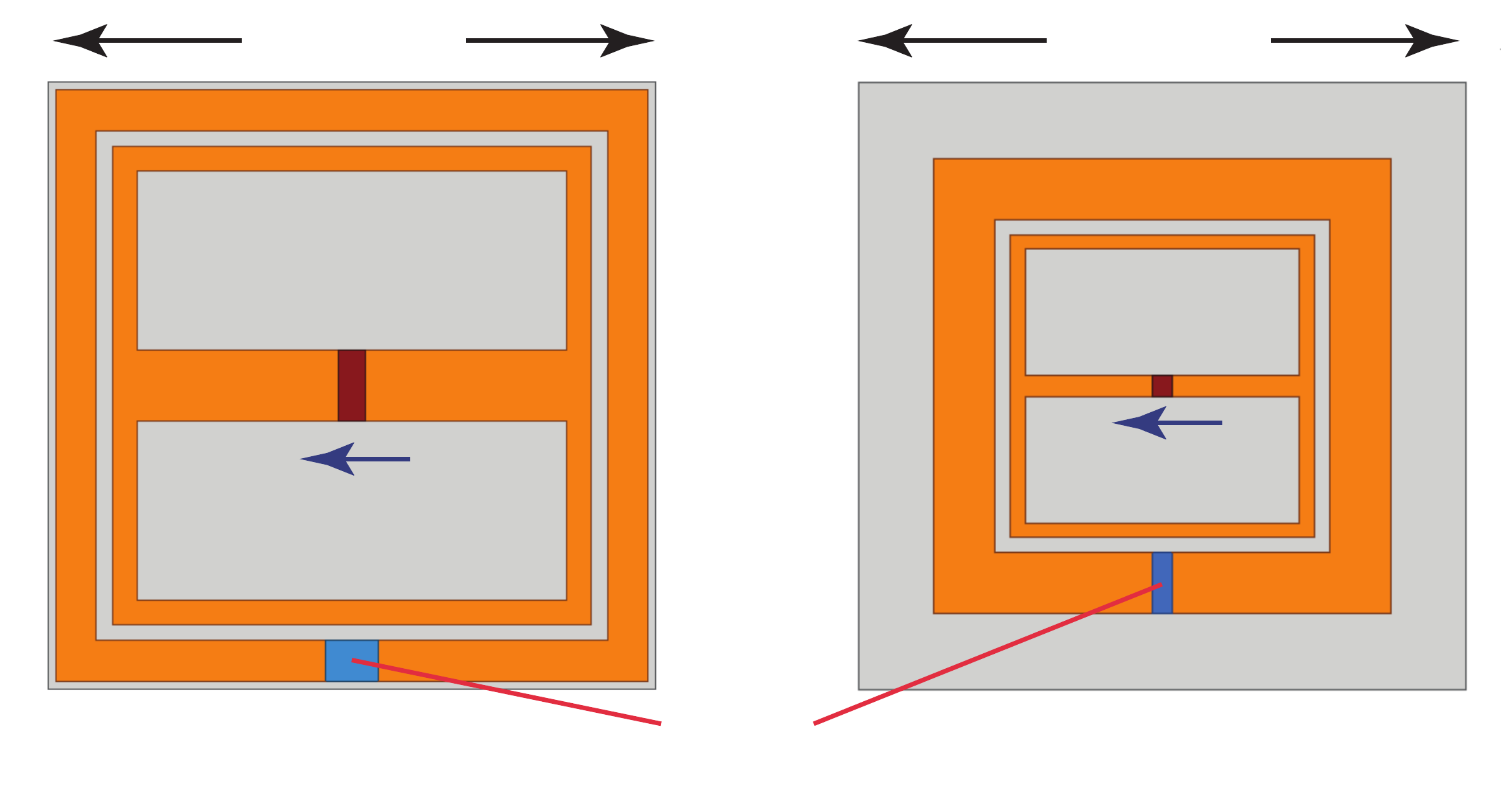}
			\put(20,51){\scriptsize $\lambda/10$}
			\put(72,51){\scriptsize $\lambda/3.33$}
			\put(23, 35){\makebox(0,0){\scriptsize \shortstack{Resistance\\ $R$}}}
			\put(77, 33){\makebox(0,0){\scriptsize \shortstack{Resistance\\ $R$}}}
			\put(50,2.5){\makebox(0,0){\scriptsize \color{amber}\boxed{\shortstack{Lumped Capacitor, $C$}}}}
			\put(24,20){\makebox(0,0){\scriptsize $E$}}
			\put(78,23){\makebox(0,0){\scriptsize $E$}}

	\end{overpic}
		\label{lambda_by_10_n_3_cells}}
	\subfigure[][]{
	\begin{overpic}[grid=false, width=0.95\columnwidth]{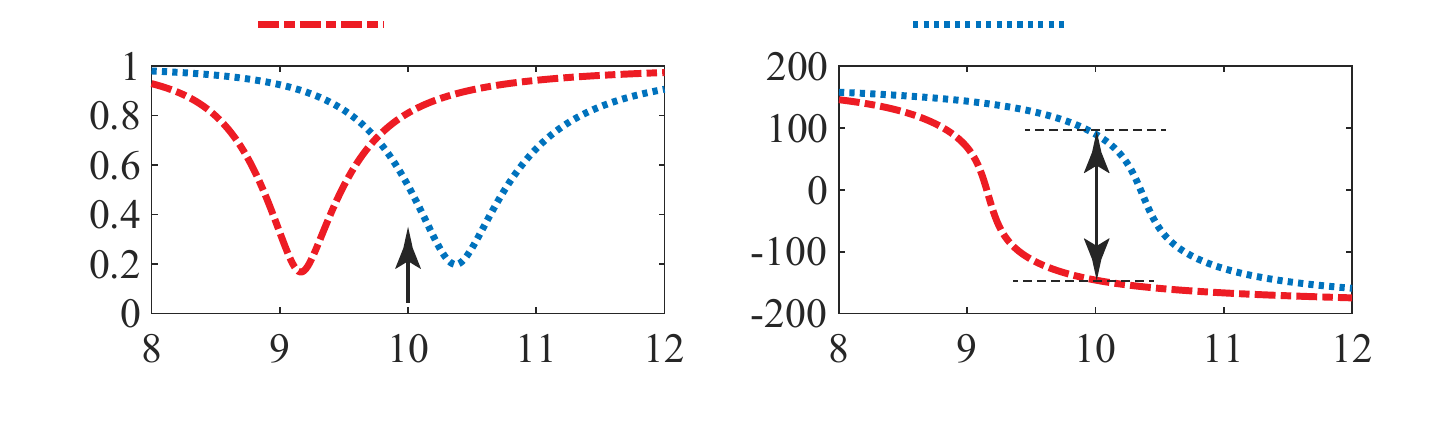}
			\put(18,3){\scriptsize Frequency (GHz)}
			\put(1,12){\rotatebox{90}{\scriptsize Magnitude}}
			\put(68,3){\scriptsize Frequency (GHz)}
			\put(49,11){\rotatebox{90}{\scriptsize Phase (deg)}}	
			\put(28.5,18){\makebox(0,0){\scriptsize $f_0$}}
			\put(72,13){\rotatebox{90}{\scriptsize $\approx 200^\circ$}}			
			\put(29,28.5){\scriptsize $C$ = 0.31~pF}
			\put(76,28.5){\scriptsize $C$ = 0.165~pF}
			\put(96,27){\rotatebox{-90}{\scriptsize \color{amber}\boxed{\shortstack{\textsc{Small Cell}\\ $\Lambda=\lambda/10$}}}}
	\end{overpic}
		\label{mag_phs_lambda_by_10_cells_cap_range}}
		
			\subfigure[][]{
	\begin{overpic}[grid=false, width=0.95\columnwidth]{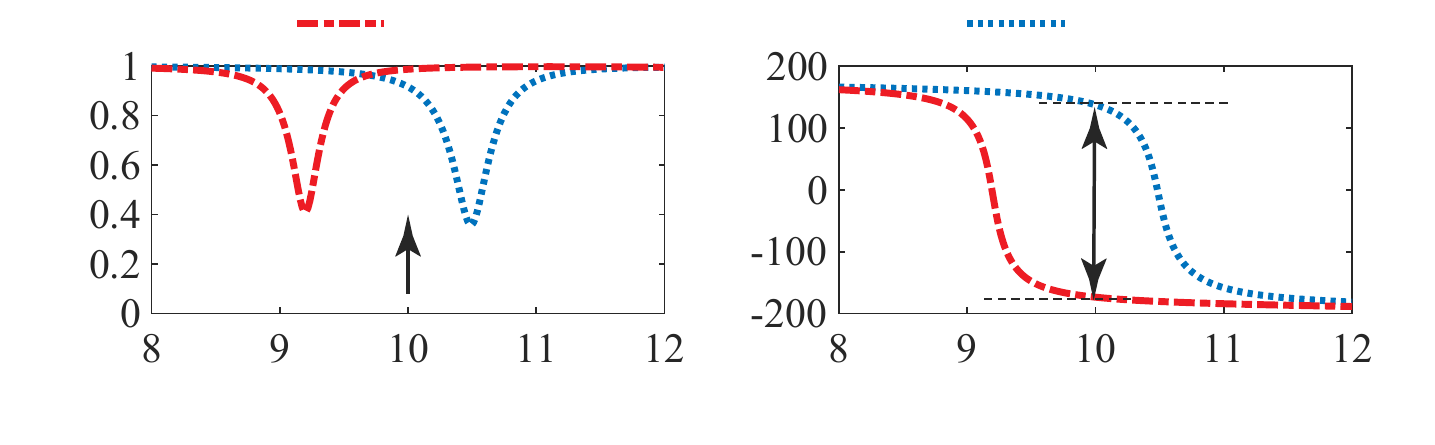}
			\put(18,3){\scriptsize Frequency (GHz)}
			\put(1,12){\rotatebox{90}{\scriptsize Magnitude}}
			\put(68,3){\scriptsize Frequency (GHz)}
			\put(49,11){\rotatebox{90}{\scriptsize Phase (deg)}}	
			\put(28.5,18){\makebox(0,0){\scriptsize $f_0$}}
			\put(72,13){\rotatebox{90}{\scriptsize $\approx 360^\circ$}}
			\put(29,28.5){\scriptsize $C$ = 2.2~pF}
			\put(76,28.5){\scriptsize $C$ = 0.3~pF}
			\put(96,27){\rotatebox{-90}{\scriptsize \color{amber}\boxed{\shortstack{\textsc{Large Cell}\\ $\Lambda=\lambda/3.33$}}}}

	\end{overpic}
		\label{mag_phs_lambda_by_3p33_cells_cap_range}}
	\caption{Impact of the unit cell sizes on the lumped capacitor ranges. (a) Schematics of the top view showing the foot-print shapes of the two unit cell sizes of $\lambda/10$ and (b) $\lambda/3.33$, resepctively (schematics are not to scale). The simulation reflection magnitude and phase to illustrate the capacitance dynamic range for two different metasurface unit cell sizes of (b) $\lambda/10$ and (c) $\lambda/3.33.$ operating at 10~GHz with a fixed lumped resistance $R = 50~\Omega$.}
	\label{mag_phs_lambda_by_10_n_3p33_cells_cap_range}
\end{figure}

It is also observed from the two-dimension contour views shown in Fig.~\ref{Amp_Phs_RC_two_resonators_ckt_hfss} that the capacitance range increases (defined around the valley region where the magnitude and phase vary the most) and moves towards larger values when operating at lower frequencies (longer wavelengths) for a \emph{fixed size unit cell} while maintaining a similar coverage of independent reflection magnitudes and phases. This observation additionally suggests that larger capacitance values for a fixed operating frequency require an electrically larger unit cell. Fig.~\ref{mag_phs_lambda_by_10_n_3p33_cells_cap_range} shows the simulated reflection magnitude and phase at a fixed resistance of $R = 50~\Omega$ for two different sizes unit cells of $\lambda/10$ and $\lambda/3.33$, when the desired operating frequency is 10~GHz. To maintain a similar magnitude and phase coverage, the capacitance values for the electrically larger unit cell ($\lambda/3.33$) is ranged between $0.3 - 2.2$~pF while it is $0.165 - 0.31$~pF for the smaller cell. Therefore, if one chooses to use higher capacitance values in the design, the unit cell may become larger based on its size with respect to the wavelength while maintaining similar magnitude-phase coverage. Such a relationship may be apparent through the cell geometries used in Fig.~\ref{mag_phs_lambda_by_10_n_3p33_cells_cap_range}, where a smaller inter-cell electromagnetic coupling due to larger cell to cell separation in $\lambda/3.33$ case, could be compensated using a larger lumped element capacitance. While this brings the flexibility of using practical varactor elements exhibiting larger lumped capacitances (thus less sensitive and economically cheaper), larger unit cell sizes may be unfavorable since they sample the required spatially varying amplitude/phase distributions poorly. Therefore, there exists a trade-off between lumped capacitor ranges and the spatial amplitude-phase discretization, where a judicious balance must be made in a practical metasurface design.

\section{Conclusion}

A novel metasurface unit cell architecture has been proposed to enable independent control of the reflection magnitude and phase at a desired operating frequency while maintaining linear polarization of the incoming fields. The proposed structure is based on a coupled-resonator configuration where a DRR loaded with PIN diode as a tunable resistive element and SRR loaded with a varactor diode as a tunable capacitor, are superimposed. The surface is next operated around one of the coupled resonant frequency, where an independent tuning of the varactor and PIN diode elements enable a wide coverage of reflection amplitude-phase, which is significantly larger than what would have been achievable using a single resonator configuration. An insightful equivalent circuit model has also been developed for investigating the amplitude-phase characteristics of a uniform surface as a function of variable resistance and capacitance. Finally, using a variety of full-wave examples, the usefulness of simultaneous and independent amplitude-phase control has been demonstrated, including cases of variable pattern gain with beam tilting and multi-beam pattern realization, which otherwise would not be possible using amplitude or phase control only.

This feature of independent phase control thus offers a practically useful mechanism to enable wave transformation which otherwise is not possible using existing conventional approaches, where either only amplitude or phase control has so far been shown. Since the proposed metasurface is based on coupled-resonators loaded with separate lumped elements, the biasing network is simple to design where external voltage controls can be separately designed with minimal inter-dependence. While this work emphasized on the inner workings and exploring the electromagnetic properties of the proposed metasurface unit cell, the practical realization with external voltage biasing controls represent a standard technique for real-time control and thus pose no fundamental issue. For instance, a practical metasurface structure can be visualized having a pair of voltage controls per unit cell, so that for a surface with $N\times N$ unit cells, $2N^2$ voltage biasing lines can be used (for pixel-by-pixel control, or $2N$ controls for a row-by-row control). Moreover, an extension to multiple polarization operations can be achieved using either a more sophisticated cell with full symmetry and an increased number of resonators or devising a super-cell where the same cell is alternately rotated by $90^\circ$. The surface thus subsequently can be interfaced with control software that may provide convenient programmatic control of the surface, where different unit cells can be independently assigned specific states, according to the desired wave transformation requirements. The proposed metasurface thus can be real-time reconfigured providing a versatile control over the fields scattered off the surface while extending their capabilities based on ML/AI may further be envisioned. Therefore, with flexible software programmable controls and enhanced reflection capabilities, the proposed metasurface may truly be called a \emph{smart reflector} with applications at the RF in the area of wireless communication, sensing, and imaging.

\bibliographystyle{ieeetran}

\bibliography{2020_MS_Mag_Phs_abbrev}

\end{document}